\documentclass[a4paper,11pt]{article}
\usepackage{jcappub} % for details on the use of the package, please see the JINST-author-manual
\usepackage{tabularx}
\usepackage{subfig}

\newenvironment{conditions*}
  {\par\vspace{\abovedisplayskip}\noindent
   \tabularx{\columnwidth}{>{$}l<{$} @{${}={}$} >{\raggedright\arraybackslash}X}}
  {\endtabularx\par\vspace{\belowdisplayskip}}

\title{\boldmath Sensitivity of BEACON to Ultra-High Energy Diffuse and Transient Neutrinos}

\collaboration{\includegraphics[height=17mm]{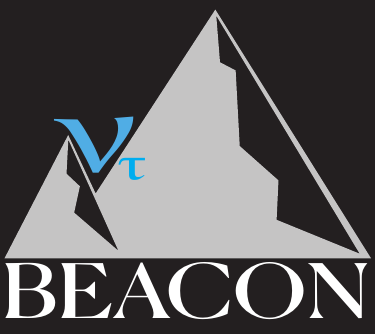}\\[6pt] BEACON Collaboration}

\author[a,b,1]{A.~Zeolla\note{Corresponding author.},}
\author[c]{J.~Alvarez-Mu\~{n}iz,}
\author[c]{S.~Cabana-Freire,}
\author[d]{W.~Carvalho~Jr.,}
\author[a,b,e]{A.~Cummings,}
\author[f]{C.~Deaconu,}
\author[a]{J.~Hinkel,}
\author[g]{K.~Hughes,}
\author[a,b]{R.~Krebs,}
\author[g]{Y.~Liu,}
\author[h]{Z.~Martin,}
\author[i, j]{K.~Mulrey,}
\author[g]{A.~Nozdrina,}
\author[f]{E.~Oberla,}
\author[k]{S.~Prohira,}
\author[l]{A.~Romero-Wolf,}
\author[f,h,m]{A.~G.~Vieregg,}
\author[a,b,e]{S.~A.~Wissel,}
\author[c]{and E.~Zas}

\affiliation[a]{Department of Physics, Pennsylvania State University, University Park, PA 16802, USA}
\affiliation[b]{Center for Multimessenger Astrophysics, Institute of Gravitation and the Cosmos, Pennsylvania State University, University Park, PA 16802, USA}
\affiliation[c]{Instituto Galego de F\'\i sica de Altas Enerx\'\i as IGFAE, Univerisade de Santiago de Compostela, 15782 Santiago de Compostela, Spain}
\affiliation[d]{Faculty of Physics, University of Warsaw, 02-093, Warsaw, Poland}
\affiliation[e]{Department of Astronomy and Astrophysics, Pennsylvania State University, University Park, PA 16802, USA}
\affiliation[f]{Department of Astronomy and Astrophysics, Kavli Institute for Cosmological Physics, University of Chicago, Chicago, IL 60637, USA}
\affiliation[g]{Department of Physics, The Ohio State University, Columbus, OH 43210, USA}
\affiliation[h]{Department of Physics, Kavli Institute for Cosmological Physics, University of Chicago, Chicago, IL 60637, USA}
\affiliation[i]{Department of Astrophysics / IMAPP, Radboud University Nijmegen, 6500 GL, Nijmegen, The Netherlands}
\affiliation[j]{NIKHEF, Science Park Amsterdam, 1098 XG, Amsterdam, The Netherlands}
\affiliation[k]{Department of Physics and Astronomy, University of Kansas, Lawrence, KS 66045, USA}
\affiliation[l]{Jet Propulsion Laboratory, California Institute for Technology, Pasadena, CA 91109, USA}
\affiliation[m]{Enrico Fermi Institute, University of Chicago, Chicago, IL 60637, USA}

\emailAdd{azeolla@psu.edu}

\abstract{Ultra-high energy neutrinos ($E>10^{17}$ eV) can provide insight into the most powerful accelerators in the universe, however their flux is extremely low. The Beamforming Elevated Array for COsmic Neutrinos (BEACON) is a detector concept which efficiently achieves sensitivity to this flux by employing phased radio arrays on mountains, which search for the radio emission of up-going extensive air showers created by Earth-skimming tau neutrinos. Here, we calculate the point-source effective area of BEACON and characterize its sensitivity to transient neutrino fluences with both short ($<15\,\text{min}$) and long ($> 1\,\text{day}$) durations. Additionally, by integrating the effective area, we provide an updated estimate of the diffuse flux sensitivity. With just 100 stations, BEACON achieves sensitivity to short-duration transients such as nearby short gamma-ray bursts. With 1000 stations, BEACON achieves a sensitivity to long-duration transients, as well as the cosmogenic flux, ten times greater than existing experiments at 1 EeV. With an efficient design optimized for ultrahigh energy neutrinos, BEACON is capable of discovering the sources of neutrinos at the highest energies.}

\begin{document}
\maketitle
\flushbottom

\section{Introduction}
\label{sec:intro}

Ultra-high energy ($E>10^{17}$ eV) neutrinos can be produced by ultra-high energy cosmic ray (UHECRs) interactions either within the source environment or during propagation \cite{Ackermann:2019cxh}. In the first instance they are known as astrophysical, while in the latter cosmogenic. Astrophysical neutrinos are excellent messengers since they can propagate extragalactic distances undisturbed, while a measurement of the cosmogenic spectrum would provide valuable information about the origin of UHECRs. %mass composition. 
Additionally, UHE neutrinos can be used to probe the fundamental physics of electroweak interactions at energies far greater than those achieved with man-made accelerators \cite{Ackermann:2022rqc}. To detect a sizable number of UHE neutrinos in a reasonable time frame however, the next-generation of detectors must have significantly larger detector volumes.

The origin of of UHECRs remains unknown in part because intervening magnetic fields deflect them during propagation. Highly luminous astrophysical transients---violent cosmic events lasting on the order of seconds to months---may contribute a significant fraction of the UHECR flux \cite{Guepin:2022qpl}. Examples of transients include blazars, magnetars, and gamma-ray bursts. Astrophysical neutrinos, produced in UHECR sources and unaffected by intervening magnetic fields, present a solution to identifying these extremely energetic point-sources. IceCube was the first to demonstrate this when it discovered a transient high-energy ($\sim10^{14}$ eV) neutrino source, blazar TXS 0506+056, associated with a gamma-ray flare in 2017 \cite{IceCube:2018cha}. More recently, IceCube discovered a point-source of neutrinos from the active galaxy NGC1068 \cite{IceCube:2022der}. In the UHE range, a 120 PeV neutrino has been detected by KM3NeT; however, it is as of yet not associated with any known sources \cite{KM3NeT:2025npi}. %An ultra-high energy neutrino point-source thus remains undiscovered. 
The UHE neutrino sky is thus largely unexplored and can provide unique insights into particle astrophysics at the highest energies.

Astrophysical sources are expected to produce mainly electron and muon neutrinos, however flavor oscillations over extragalactic distances should result in an even flavor ratio at Earth ($\nu_e : \nu_\mu : \nu_\tau=1:1:1$) \cite{Beacom}. This, combined with the increased neutrino cross section at ultra-high energies, presents a unique opportunity for detection. Tau neutrinos which skim the Earth can undergo a charged-current interaction within the crust, producing a tau lepton which then escapes into the lower atmosphere. The tau lepton can then decay, producing an up-going extensive air shower (EAS), which in turn will produce an impulsive radio signal predominantly due to the geomagnetic effect \cite{Zas_2005, Schr_der_2017}. This arrangement is unique to the tau flavor, as electrons quickly shower inside the Earth and relatively long-lived muons propagate beyond the lower atmosphere. Additionally, since electron and muon neutrinos produced by cosmic ray interactions in the atmosphere do not propagate a sufficient distance for significant flavor oscillation, any detected ultra-high energy tau neutrino is highly likely to be astrophysical in origin.

Radio (30--1000 MHz) has an in-air propagation length of $\mathcal{O}(100)$ km, enabling highly efficient, large detector volumes through the use of elevated radio antenna arrays. The Beamforming Elevated for COsmic Neutrinos (BEACON) is a novel detector concept consisting of $\mathcal{O}(100-1000)$ independent, phased radio arrays placed on mountains, which monitor a large area for the up-going EAS created by Earth-skimming tau neutrinos \cite{Wissel_2020}. Each array consists of $\mathcal{O}(10)$ low-cost, short-dipole antennas which are digitally phased together at the trigger level. Digital phasing attempts to align the waveforms received by each antenna according to pre-calculated time-delays corresponding to different arrival directions, or "beams". The aligned waveforms are then summed, improving the signal-to-noise ratio and lowering the energy threshold \cite{ROMEROWOLF201572, ALLISON2019112}. Additionally, the trigger threshold of each beam is individually tunable, allowing a higher threshold in the direction of background or a lower threshold in directions of interest. By also placing antennas outside each phased array, we can use the large interferometric baselines to achieve degree-scale pointing resolution \cite{Wissel:20248V}. % A schematic of the BEACON concept is shown in figure~\ref{fig:i}.

An initial estimate of the neutrino diffuse flux sensitivity of BEACON was presented in ref.~\cite{Wissel_2020}. Here, we present an updated estimate as well as the point-source sensitivity using an improved Monte Carlo simulation. In Section \ref{sec:sim} we describe this new simulation. In Section \ref{sec:results} we present the results of the Monte Carlo for one possible arrangement of BEACON. In Section \ref{sec:topo} we discuss how topography may affect the sensitivity estimate. Lastly, in Section \ref{sec:discussion} we discuss the implications of these results, and we conclude in Section \ref{sec:conclusion}.

\section{Monte Carlo Simulation}
\label{sec:sim}

The effective area of an observatory to point sources of neutrinos $A(t, E_\nu,\hat{r}_i)$, derived in \cite{ANITA:2021xxh}, is given by 

\begin{equation}
   A(t,E_\nu,\hat{r}) = \int_{A_g} dA_g \:(\hat{r} \cdot \hat{u}) \:\Theta(\hat{r} \cdot \hat{u}) \: P_\text{obs}(t, E_\nu,\hat{r}, \hat{u}) \label{eq:eff_area}
\end{equation}
where:
\begin{conditions*}
t & the observation time, \\
E_\nu & the neutrino energy, \\
\hat{r} & the vector extending from the point source to Earth, \\
A_g & the geometric area on the surface of the Earth that is integrated over, \\
\hat{u} & the location of the differential area on the surface of the Earth, \\
\Theta & the Heaviside step-function, \\
P_\text{obs}(t, E_\nu, \hat{r}, \hat{u}) & the probability of detection.
\end{conditions*}

\begin{figure}[tbp]
\centering
\includegraphics[width=\textwidth]{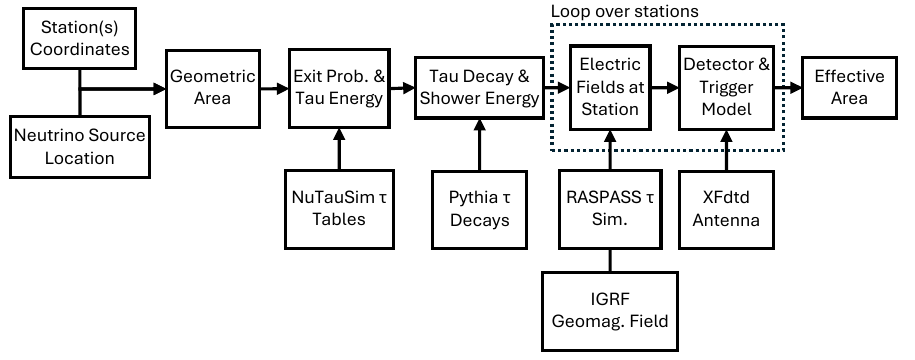}
\caption{Flowchart depicting the MARMOTS Monte Carlo procedure. Packages used in the Monte Carlo include NuTauSim~\cite{NuTauSim, Alvarez-Muniz:2018owm} to sample neutrino interactions in the Earth, Pythia~\cite{PYTHIA} for the tau decays in air, ZHAireS-RASPASS~\cite{Tueros:2023mxa, Tueros:2024bzl} and the IGRF \cite{IGRF} for the radio emission of air showers, and the electromagnetic simulation package XFdtd~\cite{xfdtd} for the antenna response.}
\label{fig:flow} 
\end{figure}

We developed a Monte Carlo, named the \textbf{M}ultiple \textbf{A}ntenna A\textbf{r}rays on \textbf{Mo}untains \textbf{T}au \textbf{S}imulation, or MARMOTS, to numerically solve Eq.~\ref{eq:eff_area} and determine the point-source effective area of BEACON. MARMOTS can calculate the effective area of any number and configuration of elevated phased arrays, an important feature since the layout and location of BEACON has not yet been finalized. MARMOTS is open source and available on GitHub\footnote{https://github.com/beaconTau/marmots}. The structure of MARMOTS is based on Tapioca, the point-source effective area model for ANITA \cite{ANITA:2021xxh}. ANITA was a balloon-borne neutrino observatory which observed the Antarctic ice from high altitude. Here, we must account for many stationary phased arrays which can be distributed globally, rather than a single moving array. 

MARMOTS begins by dividing the sky into pixels of equal solid angle using HEALPix \cite{Zonca2019, 2005ApJ...622..759G}. Each pixel is then a point-source location ($-\hat{r}$) expressed in right ascension ($\alpha$) and declination ($\delta$). Given the desired location of each phased array, the Monte Carlo then calculates the geometric area ($A_g$) containing potentially detectable exiting tau leptons associated with each point-source. Tau lepton exit points are then uniformly generated in this area. $P_\text{obs}$ is then determined for each of these tau leptons by splitting up the calculation into two components: the probability of the tau lepton exiting the earth ($P_\text{exit}$), and the probability that the resulting shower is subsequently detected by any of the arrays ($P_\text{detect}$). The point-source effective area in the direction of each source is thus given by
\begin{equation}
   A(E_\nu, \alpha, \delta)  = 
   \frac{A_g}{N} \sum_{i=1}^{N} (\hat{r}_i \cdot \hat{u}_{i})\:\Theta(\hat{r}_i \cdot \hat{u}_i) \: P_{i,\text{exit}} \: P_{i,\text{detect}}, \label{eq:sum}
\end{equation}
where $N$ is the number of exiting tau leptons generated within area $A_g$.

A flow-chart depicting the procedure used by MARMOTS is shown in figure \ref{fig:flow}. We will now discuss in more detail the geometry calculation ($A_g, \, \hat{r}_i, \, \hat{u}_{i}$), tau lepton exit probability calculation ($P_\text{exit}$), and the radio emission and detector models ($P_\text{detect}$) used in MARMOTS.

\subsection{Geometry}
\label{sec:geom}

\begin{figure}[tbp]
 \centering
 \subfloat{{\includegraphics[width=0.58\textwidth]{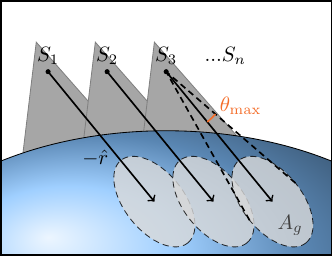}}}
 \,
 \subfloat{{\includegraphics[width=0.40\textwidth]{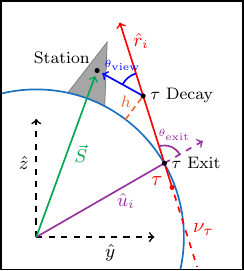}}}
  \caption{Left: Diagram demonstrating the calculation of $A_g$ in MARMOTS. From each station ($S_n$), a cone with opening angle $\theta_\text{max}$ is extended in the direction of the source ($-\hat{r}_i$). The intersection of each cone with the Earth is found, and the union of these intersection areas is $A_g$. Right: Relevant geometry for a single event, whose exit point is sampled from $A_g$. The dot-product of $\hat{r}_i$ and $\hat{u}_i$ is used in the geometric area calculation, while $\theta_\text{exit}$, $\theta_\text{view}$, and $h$ are used in the calculation of $P_\text{exit}$ and $P_\text{detect}$.}
  \label{fig:geometry}
\end{figure}

MARMOTS models the Earth as a smooth sphere. An arbitrary number of stations can be defined by their latitude, longitude, and altitude. One can also specify the orientation of each station relative to geographic East, the azimuthal field-of-view, and the number of phased antennas. Each station is then modeled as a single point at its specified location, floating above a smooth Earth. An alternate version of MARMOTS, which includes local topography, is discussed in section \ref{sec:topo}.

For each source location ($\alpha$, $\delta$), the geometric area containing detectable exiting tau leptons $A_g$ is calculated. This is done by extending a cone with opening angle $\theta_\text{max}$ from each station in the direction of the source ($-\hat{r}$). The intersection between the cone and the surface of the Earth, a curved ellipse, is the geometric area containing detectable exiting tau leptons for a single station. For mountaintop geometries, we use a $\theta_\text{max}$ of $3^\circ$ as exiting tau leptons at greater angles are unlikely to be detected \cite{Wissel_2020}. For some source directions, a station may not view the Earth at all. In this instance, the intersection area for that station is zero and it is excluded from further simulation. The overall geometric area $A_g$ is the union of all the individual intersection areas. This geometry is shown schematically on the left in figure \ref{fig:geometry}.

The union of polygons defined on the surface of a sphere is difficult to calculate, so the intersection areas are first projected onto a 2D plane. This is done via sinusoidal projection which conserves area \cite{snyder}. The union of the now 2D polygons is found using the Python package Shapely \cite{shapely2007}. From Shapely we also obtain the area of the union ($A_g$) in square-kilometers. The union area is then triangulated using a Delaunay triangulation algorithm \cite{shewchuk96b}. This allows us to uniformly sample tau lepton exit points within the arbitrarily-shaped area in an efficient manner. The exit points are then inverse sinusoidally projected back onto a sphere. The result is the latitude and longitude of $N$ tau lepton exit points uniformly generated within $A_g$. The source is approximated as being infinitely far away resulting in parallel propagation axes. The propagation axis of each tau lepton ($\hat{r}_i$) is therefore equal to $\hat{r}$. The dot product ($\hat{r}_i \cdot \hat{u}_{i}$) is then calculated, where $\hat{u}_{i}$ is a vector from the center of Earth to each exit point, as shown on the right in figure \ref{fig:geometry}. A negative dot product indicates that the tau lepton is propagating into the Earth and is thus undetectable. These events are excluded from further simulation (this is the Heaviside step function in eq. \ref{eq:sum}).

\begin{figure}[tbp]
  \centering
  \includegraphics[width=1.0\textwidth]{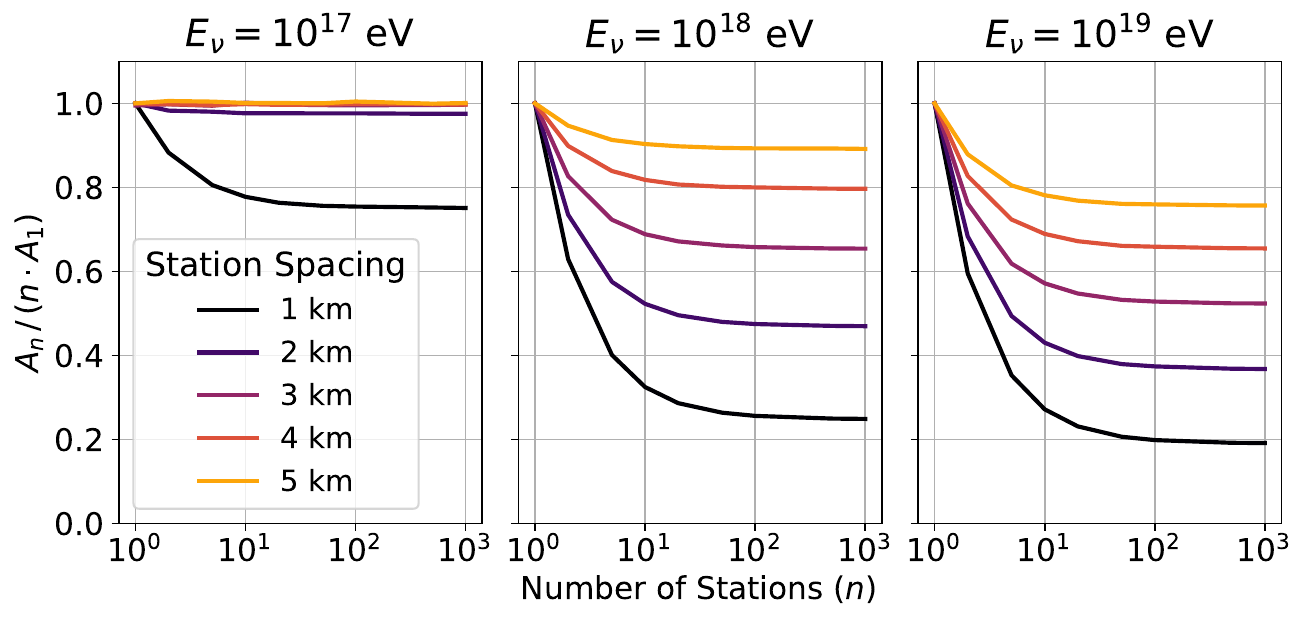}
  \caption{The ratio of the effective area of $n$ stations ($A_n$) to the effective area of a single station ($A_1$) multiplied by $n$, as a function of $n$. Shown for a variety of station spacings and for three different neutrino energies. The stations are arranged in a line with the neutrino source located at an azimuth angle of $0^\circ$ and a zenith angle of $93^\circ$. $A_n/(n \cdot A_1) =1$ if individual effective areas do not overlap. As expected, stations spaced farther apart have reduced overlap. Additionally, the effect is energy dependent, with greater overlap occurring at higher energies.}
  \label{fig:overlap}
\end{figure}

This method, which finds the union of the individual geometric areas, properly accounts for potentially overlapping effective areas. In the previous sensitivity estimate for BEACON, it was assumed that the stations would be sufficiently spread apart such that their individual effective areas would not overlap, and thus the effective area of $n$ stations ($A_n$) would simply be the effective of a single station ($A_1$) multiplied by $n$ \cite{Wissel_2020}. When overlap is present however, the total effective area will be less than this. The effect of overlapping individual effective areas on the total effective area is demonstrated in figure \ref{fig:overlap}. No overlap corresponds to $A_n/(n \cdot A_1) =1$. As expected, stations spaced farther apart have reduced overlap between them. Additionally, the effect is energy dependent. Higher energy events can be detected farther off-axis, requiring the stations to be spaced farther apart to view independent neutrino interaction volumes. For $E_\nu=10^{17}$ eV, the stations are nearly independent at spacings of 2 km, while at higher energies the stations must be spaced farther than 5 km apart in order to be truly independent. 

We also see that the non-linearity of the overlap effect eventually subsides as we continue to add stations. This is likely due to the fact that as the line of stations becomes longer, overlaps only continue to occur between stations and their nearby neighbors. We find that, for this particular arrangement, the effective area of 1000 stations is $\sim10$ times higher than the effective area of 100 stations regardless of spacing and neutrino energy. While overlap reduces the total effective area, it is difficult to avoid entirely due to potential size constraints. One upside to overlapping effective areas is improved event reconstruction, as events will be viewed with additional interferometric baselines.

\subsection{Tau Lepton Exit Probability and Decay}
\label{sec:exit}

MARMOTS uses the same models of tau lepton exit and decay as our previous simulation \cite{Wissel_2020} and Tapioca \cite{ANITA:2021xxh}. Lookup tables, generated using NuTauSim \cite{NuTauSim, Alvarez-Muniz:2018owm}, contain exiting tau lepton energies and exit probabilities as a function of exit angle and incident neutrino energy. For each exiting tau lepton generated within $A_g$, the exit angle ($\theta_\text{exit}$ in figure \ref{fig:geometry}) is calculated. The tau lepton energy and exit probability ($P_{i,\text{exit}}$) is then sampled from the appropriate NuTauSim distribution. 

The main source of uncertainty in the effective area calculation appears in this calculation of the exit probability since the neutrino cross section and the tau lepton energy loss are needed in an energy range in which experimental data does not yet exist. It is not the purpose of this work to perform a thorough study of the effect of these uncertainties, however, for ANITA, a study performed in \cite{Romero-Wolf:2018zxt} concluded that the uncertainty in neutrino interaction cross sections had a small effect on the estimated exposure to tau neutrinos below $10^{20}$ eV. This was because increasing the cross section decreased $P_\text{exit}$ at steep emergence angles, but increased $P_\text{exit}$ at shallow emergence angles. When decreasing the cross section, the opposite was instead true. For BEACON, which operates at a much lower elevation than ANITA, the typical emergence angles are smaller, around $\sim 2^\circ$. At these angles, the combined uncertainties in the neutrino interaction cross section and tau energy loss can lead to a factor of $\sim 2$ variation in $P_{\text{exit}}$ at $10^{20}$ eV, as shown in Fig. 15 of \cite{NuTauSim, Alvarez-Muniz:2018owm}. Further studies are needed to validate these expectations.

The distance each exiting tau lepton propagates before decaying is randomly sampled from an exponential distribution given the tau lepton lifetime and energy. The fraction of energy transferred from each tau lepton to the resulting extensive air shower is determined by randomly sampling from a distribution generated using PYTHIA \cite{PYTHIA}. EAS initiated by muon decay are highly unlikely to be detected, given the large muon decay length at 100 PeV ($6.25 \times 10^8$ km). EAS can also be initiated by muon-air interactions, however most significant energy depositions would occur above the altitude of BEACON. This is due to the muon-air interaction length for an energy deposition of at least 10$\%$ being $\sim 10^5$ g/cm$^2$ (see Fig. 5 in \cite{Cummings:2020ycz}), which is larger than the typical grammage from the surface of the Earth to the BEACON detector. Future simulations using NuLeptonSim \cite{NuLeptonSim} will explore what percent of these showers remain detectable. For now, decays such as $\tau \rightarrow \mu \, \nu_\mu \, \nu_\tau$ ($\Gamma\sim 17.4\%$) are assumed to be undetectable, and thus return a energy fraction of zero.

\subsection{Radio Emission}

Our previous Monte Carlo used electric field look-up tables generated using ZHAireS \cite{zhaires}. We have now updated those look-up tables using ZHAireS-RASPASS \cite{Tueros:2023mxa, Tueros:2024bzl}. These new simulations properly account for a spherical ground and atmosphere, rather than using a simple flat geometry. While a flat-Earth model is sufficient for simulating air showers with shallow zenith angles, it becomes inaccurate for showers with steep zenith angles such as those targeted by BEACON. 

ZHAireS-RASPASS models atmospheric refraction using an exponential approximation and assumes straight-line propagation. A dedicated study was performed to compare the arrival times of pulses at the observer for nearly horizontal trajectories. The comparison included both straight and curved ray paths, and two atmospheric models: an exponential approximation and one where the refractive index is proportional to the atmospheric density, derived from GDAS data at the BEACON site. The differences in arrival times were found to be smaller than 0.2 ns for travel distances as large as 100 km. Such differences are expected to be irrelevant in the 30–80 MHz frequency band used by the BEACON antennas. 

The simulation parameters used to create the look-table tables remain the same. A line of detectors is placed at a specified altitude such that the detectors are uniformly spaced in view angle, $\theta_\text{view}$. Each detector observes the time-domain electric field generated by a $10^{17}$ eV extensive air shower resulting from a $\tau$-lepton decaying via its dominant decay mode ($\tau\rightarrow \nu_\tau\pi^0\pi^-$). The look-up tables store the peak electric field within 10 MHz sub-bands as a function of exit angle ($\theta_\text{exit}$), decay altitude ($h$), and view angle ($\theta_\text{view}$) (see figure \ref{fig:geometry}). The range of parameters used are the same as those in Table 1 of \cite{Wissel_2020}. The peak electric field $E_\text{peak}$ within a 10 MHz sub-band, with center frequency $f_c$, is found by band-pass filtering the time-domain electric fields produced by ZHAireS-RASPASS. The peak electric field across a chosen bandwidth can be approximated by summing these sub-band peaks. This approximation assumes the different sub-bands are in phase with each other, which is valid for narrow bandwidths (such as 30-80 MHz).

\begin{figure}[tbp]
  \centering
  \includegraphics[width=1.0\textwidth]{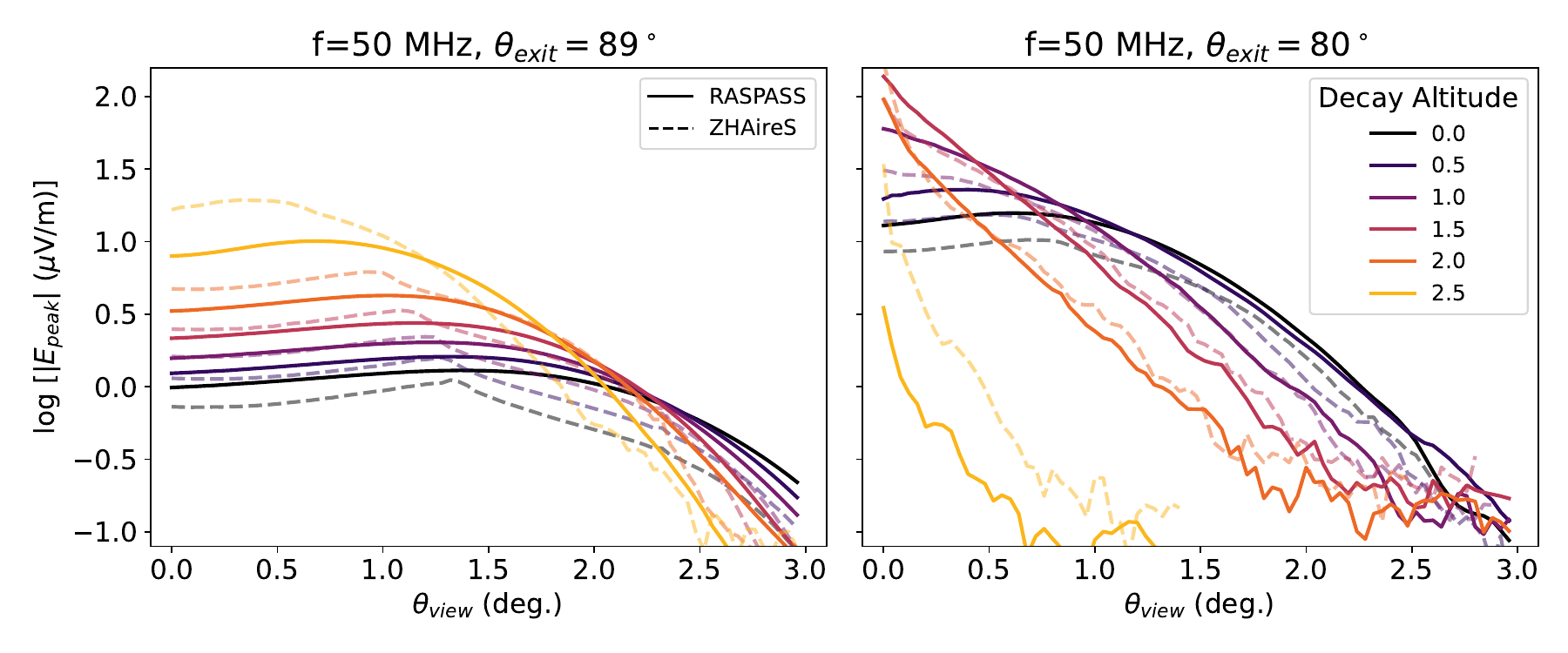}
  \caption{Peak electric field at 50 MHz, as a function of view angle, for a $10^{17}$ eV $\tau$-lepton induced extensive air shower with varying decay altitudes, at a fixed detector altitude of 3 km and exit angles of $89^\circ$ (left) and $80^\circ$ (right). The electric fields computed by RASPASS (solid) are compared to those computed by ZHAireS (dashed) in our previous study \cite{Wissel_2020}.}
  \label{fig:epeak}
\end{figure}

In figure \ref{fig:epeak}, $E_\text{peak}$(50 MHz) as a function of $\theta_{view}$ is shown for varying decay altitudes at two different exit angles ($89^\circ$ and $80^\circ$), for a detector at an altitude of 3 km. We compare $E_\text{peak}$ as determined by ZHAireS in our previous work (dashed lines) to those determined by ZHAireS-RASPASS (solid lines). Ultimately, this change seems to increase the effective area by a factor of $\sim2$. This is likely driven by the increase to $E_\text{peak}$ at low decay altitudes which, due to the exponential sampling of decay lengths, is where the majority of tau decays take place.

For each shower simulated in MARMOTS, $E_\text{interp}(f_c)$ is determined by a linear interpolation of the appropriate look-up table given the exit angle, decay altitude, and view angle. $E_\text{interp}(f_c)$ must then be scaled to account for the shower energy ($\mathcal{E})$, distance to the decay ($d$), and the differing geomagnetic field ($\vec{B}$), both its magnitude and angle with the shower axis ($\hat{v}$). MARMOTS determines the geomagnetic field at the location of the detector using the IGRF database \cite{IGRF}. Each of these effects scale the resulting electric field approximately linearly \cite{Zilles:2018kwq} so,

\begin{equation}
   E_\text{peak}(f_c)=E_\text{interp}(f_c)\cdot \frac{\mathcal{E}}{\mathcal{E}_\text{LUT}} \cdot \frac{d_\text{LUT}}{d} \cdot \frac{|\hat{v}\times\vec{B}|}{|\hat{v}_\text{LUT}\times\vec{B}_\text{LUT}|}. \label{eq:scaling}
\end{equation}

\subsection{Detector Model}

The peak power incident on each antenna, $P_\text{inc}$, is equal to $|E_\text{inc}|^2/\eta$, where $E_\text{inc}$ is the incident electric field and $\eta$ is the impedance of free space, 377 $\Omega$. The power received is equal to $A_{\text{eff}}P_\text{inc}$, where $A_{\text{eff}}$ is the effective area of the antenna. The effective area of the antenna is equal to $\lambda^2G(\theta,\phi)/4\pi$, where $\lambda$ is the wavelength and $G(\theta,\phi)$ is the gain of the antenna as a function of zenith and azimuth angle. The open circuit voltage measured at the feed of the antenna is then given by

\begin{equation}
    V_{o.c.}=\frac{2c}{f}\sqrt{\frac{G(\theta,\phi)}{4\pi}\frac{R_A}{\eta}}|E_\text{inc}|,
    \label{eq:open-circuit}
\end{equation}
where $c$ is the speed of light \cite{pozar, orfanidis}.

For a receiving antenna, the readout amplifier appears as a load being fed by a generator, the antenna. The load has impedance $Z_L$ (typically 50 $\Omega$), while the antenna has internal impedance $Z_A$, forming a voltage divider. The resistive components of $Z_L$ and $Z_A$ are denoted $R_L$ and $R_A$ respectively. The short-dipoles used by the BEACON prototype are connected to a transformer to improve the impedance matching. If the impedance transformation ratio is given by $r$, then the voltage at the readout is given by \cite{Abreu_2012}
\begin{equation}
    V_{L}=\frac{2c}{f}\left|\frac{1}{\sqrt{r}}\frac{r Z_L}{Z_A+r Z_L}\right|\sqrt{\frac{G(\theta,\phi)}{4\pi}\frac{R_A}{\eta}} \:|E_\text{inc}|.
    \label{eq:voltage}
\end{equation}
The BEACON prototype, for example, uses a 4:1 transformer and so has $r=4$.

The peak voltage in a given 10-MHz subband with center frequency $f_c$ is thus given by
\begin{equation}
    V_\text{sub}(f_c)=\frac{2c}{f_c}\left|\frac{1}{\sqrt{r}}\frac{r Z_L}{Z_A(f_c)+r Z_L}\right|\sqrt{\frac{G(f_c,\theta,\phi)}{4\pi}\frac{R_A(f_c)}{\eta}}\:E_\text{peak}(f_c).
    \label{eq:peak_voltage}
\end{equation}
The full band voltage $V_\text{peak}$ is found by summing $V_{sub}(f_c)$ across the detector's bandwidth.

The noise at each receiver is due to thermal noise, resulting from both the antenna temperature and the internal system temperature. The antenna temperature results from the sum of galactic and extragalactic noise, $T_\text{sky}$, and thermal noise due to the ground, $T_\text{ground}$. We use the Dulk parametrization to model the frequency dependence of galactic and extragalactic noise \cite{Dulk}. We assume a typical ground noise $T_\text{ground} = 300$ K. The antenna temperature depends on what fraction, $s$, of the antenna's field-of-view (FoV) is sky versus ground. The internal noise temperature $T_L$, due primarily to the low-noise amplifiers, is $T_L=100$ K. The RMS voltage is thus given by
\begin{equation}
   V_\text{RMS}= \sqrt{\int_{f_\text{low}}^{f_\text{high}}
   \left\{
   \left|\frac{1}{\sqrt{r}}\frac{r Z_L}{Z_A+r Z_L}\right| 4 k_B R_A [s T_\text{sky} + (1-s)T_\text{ground}] + k_B R_L T_L \: 
   \right\} \, df},
    \label{eq:noise}
\end{equation}
where $k_B$ is the Boltzmann constant.

The values of the gain ($G$) and impedance ($Z_A, Z_L$) can either be specified or input from prior-run antenna simulations. For BEACON, we use XFdtd \cite{xfdtd} simulations of the horizontally-polarized BEACON prototype antenna. The BEACON prototype antenna is an electrically-short dipole elevated 4 meters above a sloped ground, with a bandwidth of 30 to 80 MHz. The XFdtd antenna simulation is discussed in detail in \cite{Southall:2022yil}. Using this model, and a sky fraction of $s=0.5$, the $V_\text{RMS}$ of a single BEACON prototype antenna is 3.8 $\mu$V.

The signal-to-noise ratio (SNR) of an event for a single antenna is given by $V_\text{peak}/V_\text{RMS}$. When phasing, the signal adds together coherently. The signal thus increases by a factor of $N_A$, the number of antennas. Noise adds together incoherently, and thus increases by a factor of $\sqrt{N_A}$. The SNR of a phased array containing $N_A$ antennas is thus given by
\begin{equation}
   \text{SNR} = \sqrt{N_A}\left(\frac{V_\text{peak}}{V_\text{RMS}}\right).
    \label{eq:SNR}
\end{equation}

MARMOTS allows the user to specify the number of phased antennas in each station. If the SNR exceeds some input threshold, then a trigger occurs, and the station has detected the event. MARMOTS checks for triggers for all $N$ air showers at each of the stations. The result is an array of size $N$ containing true/false (triggered/untriggered) boolean values for each station. Ultimately, only one station needs to trigger in order for an event to have been detected. $P_\text{detect}$ is thus the union of these boolean arrays. We now have all components of equation \ref{eq:sum}.

\section{Results}
\label{sec:results}
The full-size BEACON experiment is expected to consist of $\mathcal{O}(100-1000)$ stations. Here, we simulate 100 stations, each consisting of 10 phased antennas. 10 antennas was chosen as it was found to provide an optimal balance between gain and array size in our previous work \cite{Wissel_2020}. The stations are arranged along the same line of longitude, centered on the location of the BEACON prototype (latitude: $37.589^\circ$ N, longitude: $118.238^\circ$ W) \cite{Southall:2022yil}. Each station is placed at an altitude of 3 km. The stations are spaced 3 km apart, a compromise between reducing the impact of overlap while also minimizing overall size. The stations face East to maximize the geomagnetic effect. We assume each station has a field-of-view of $120^\circ$ due to the presence of the mountain behind each one. Lastly, the SNR threshold for triggering is assumed to be 5, based on the phased-array trigger thresholds achieved by the ARA experiment \cite{Allison:2018ynt}.

\begin{figure}[tbp]
  \centering
  \includegraphics[width=0.8\textwidth]{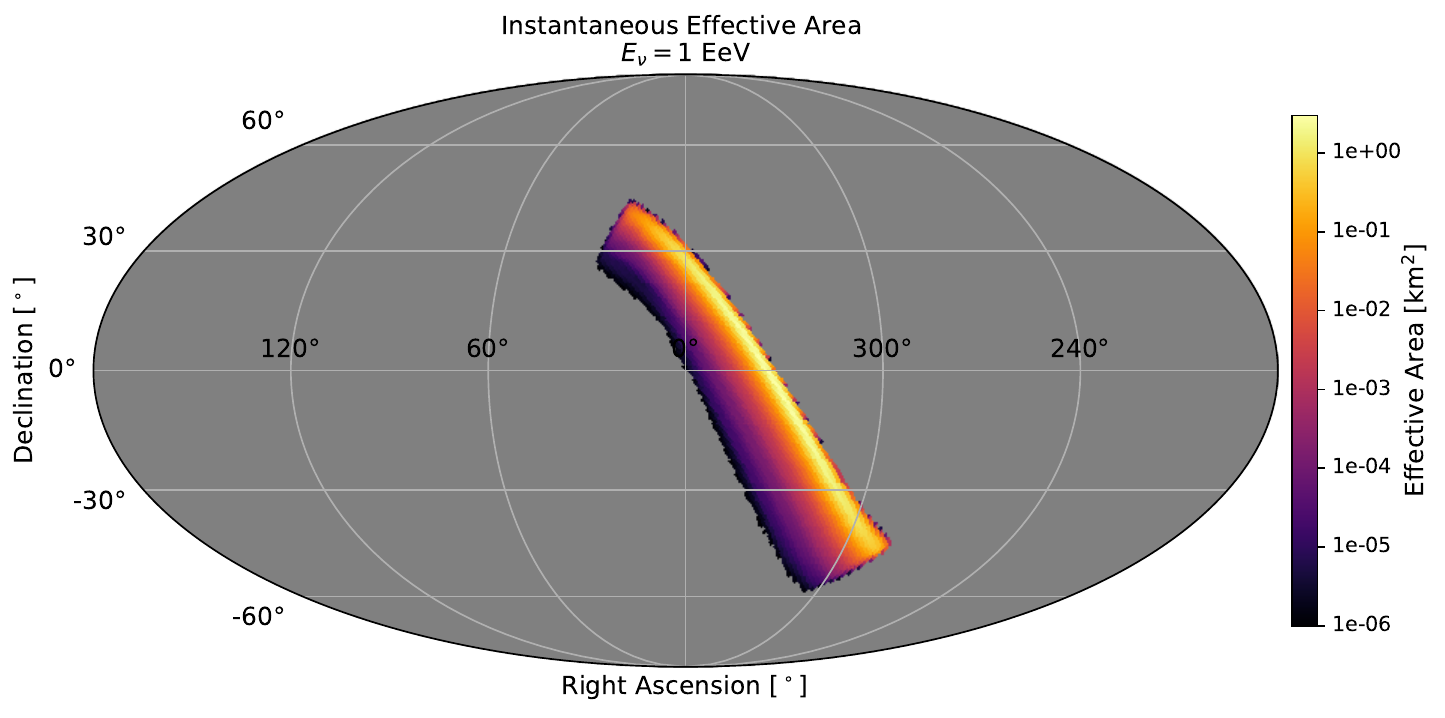}
  \caption{Mollweide projection of the instantaneous effective area to $10^{18}$ eV tau neutrinos for a 100-station BEACON as a function of source right-ascension and declination (Greenwich Sidereal Time of 0:00).}
  \label{fig:effective_area}
\end{figure}

Figure \ref{fig:effective_area} shows a Mollweide projection of the effective area calculated by MARMOTS for this simulation setup, for $E_\nu=10^{18}$ eV, at one instantaneous moment of time (Greenwich Sidereal Time of 0:00). We see that BEACON has a large effective area along a narrow band in the sky. This band corresponds to sources which lie just below the horizon. In this direction, each station views a large geometric area while simultaneously the tau lepton exit probability is maximized. As we look further below the horizon, the geometric area and exit probability simultaneously decrease and the effective area rapidly decreases. Above the horizon there is zero effective area. The instantaneous effective area at additional neutrino energies is shown in appendix \ref{sec:appendix}.

\begin{figure}[tbp]
  \centering
  \includegraphics[width=0.8\textwidth]{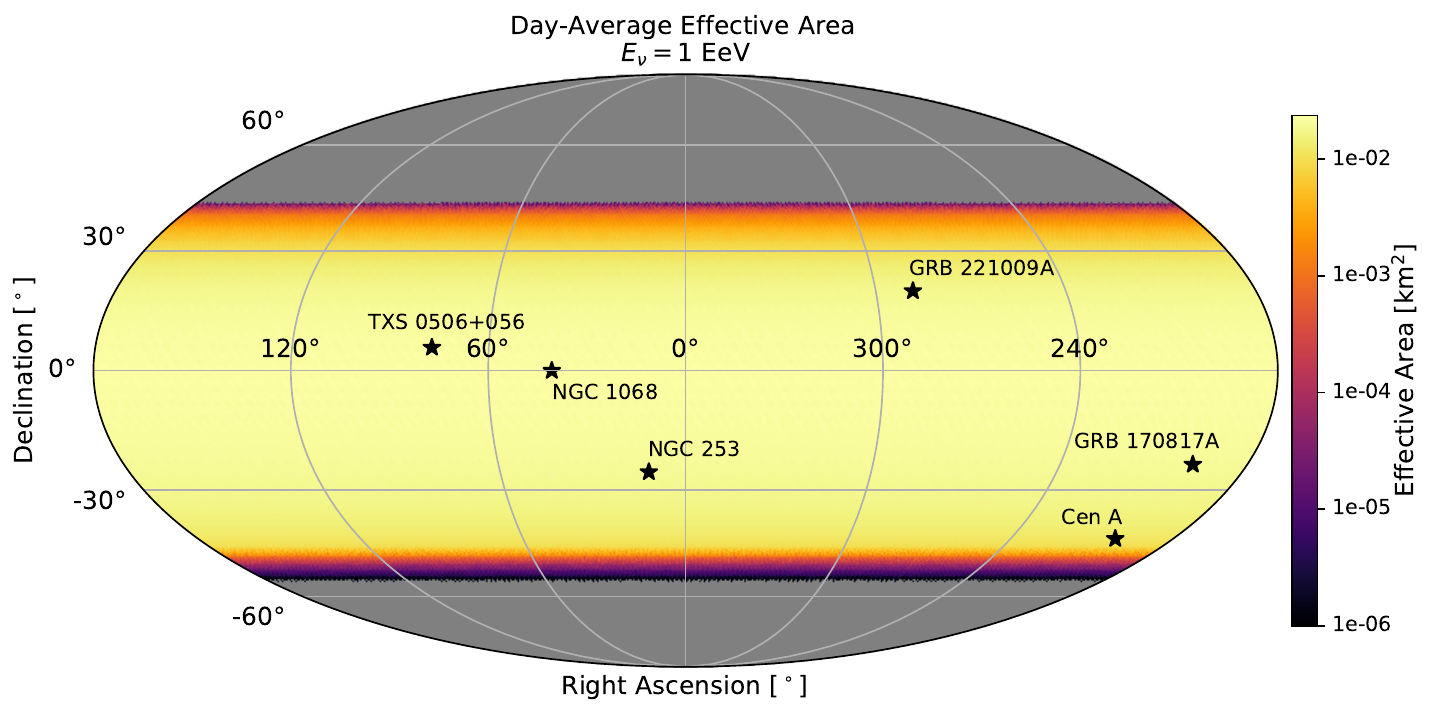}
  \caption{Mollweide projection of the day-average effective area to $10^{18}$ eV tau neutrinos for a 100-station BEACON as a function of source right-ascension and declination. Also shown are various promising ultra-high neutrino sources, past and present.}
  \label{fig:day_average}
\end{figure}

As the Earth rotates, the instantaneous effective area shown in figure \ref{fig:effective_area} will be translated in right ascension. Figure \ref{fig:day_average} shows the effective area of BEACON averaged over one day. Over the course of a day, BEACON will view $\sim70\%$ of the sky due to its large azimuthal field-of-view, Eastward facing, and position near the Equator. Overlaid the day-average effective area in figure \ref{fig:day_average} are various targets of interest (past and present) such as sources detected by IceCube (TXS 0506+056, NGC 1068) \cite{IceCube:2018cha, IceCube:2022der}, the first GRB detected via both light and gravitational waves (GRB 170817A) \cite{LIGOScientific:2017zic}, the brightest GRB of all time (GRB 221009A) \cite{Williams:2023sfk}, and nearby starburst galaxies (NGC 253, Centaurus A) \cite{Yoast-Hull:2013qfa, Mbarek:2024nvv}.

In the following sensitivity results, the instantaneous and day-average effective area of 1000 stations is estimated by scaling the 100-station effective areas by a factor of 10. This is a valid approximation since the scaling between 100 and 1000 stations is linear at all energies for a station spacing of 3 km (see figure \ref{fig:overlap}).

\subsection{Sensitivity to Point-Source Fluences}

The sensitivity of a detector to an all-flavor $dN_\nu/d{E}_\nu \propto E_\nu^{-2}$ neutrino fluence from a point-source is given by
\begin{equation}
   E_\nu^2 \phi_\nu  = 
   \frac{2.44} {\text{ln}(10)} \frac{3 \: E_\nu}{A(E_\nu)}, \label{eq:sens}
\end{equation}
where 2.44 is the Feldman-Cousins factor for a 90\% unified confidence level in which zero candidate and background events are observed \cite{Feldman:1997qc}, $\text{ln}(10)$ arises when integrating over one decade in energy, and the factor of 3 accounts for the fact that BEACON is sensitive to only one of three neutrino flavors.

\begin{figure}[tbp]
  \centering
  \includegraphics[width=0.75\textwidth]{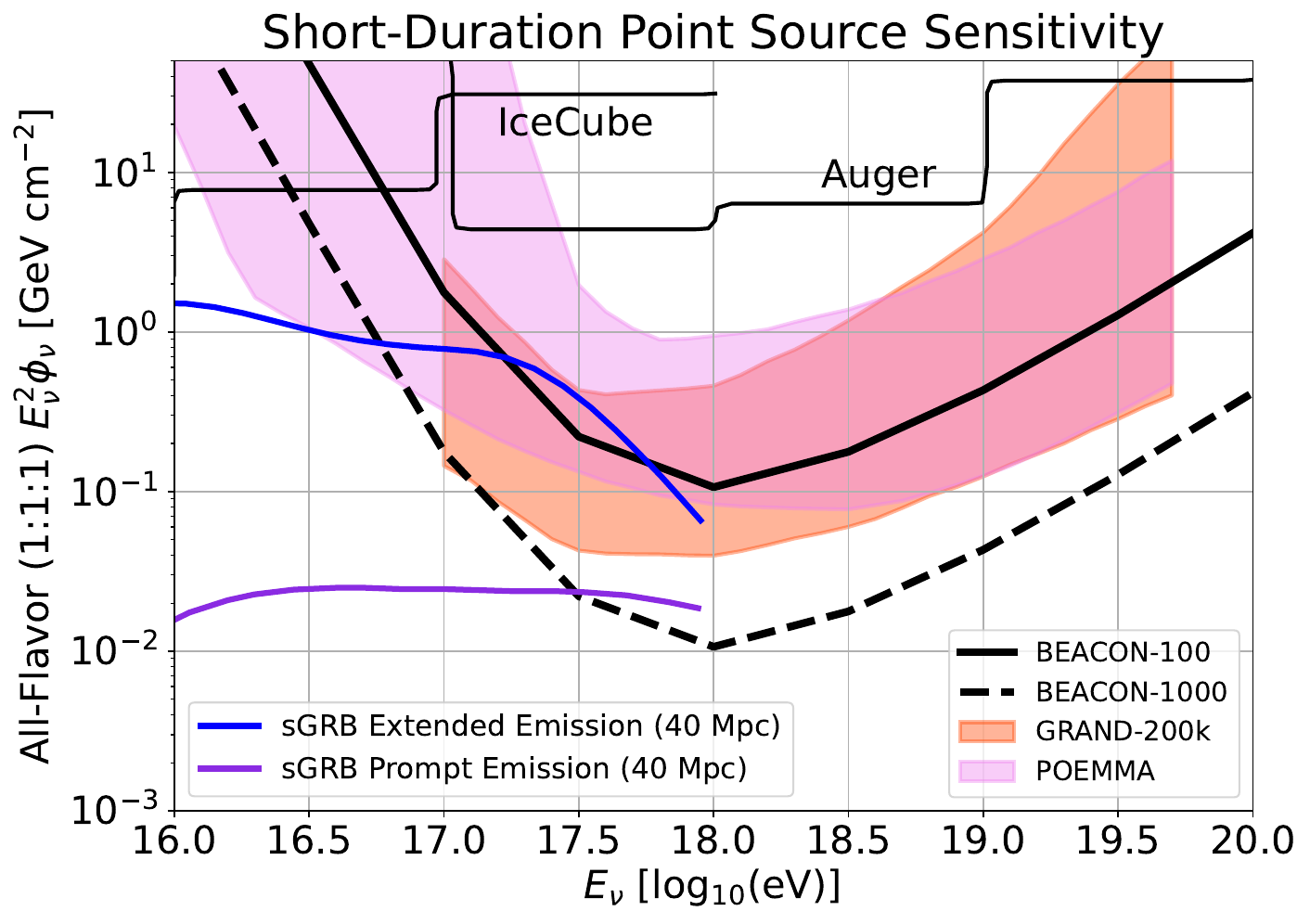}
  \caption{The sensitivity of BEACON to a short-duration ($t<1000 \, \text{s}$) neutrino fluence from a point-source. The result for 100 stations (black, solid) is obtained by inserting the maximum instantaneous effective area at each energy into eq. \ref{eq:sens}. The result for 1000 stations (black, dashed) is predicted by scaling the 100 station result by a factor of 10. These curves are compared to a model of the all-flavor fluence resulting from a short gamma ray burst, as well as short-duration limits imposed by IceCube and the Pierre Auger Observatory (see text) \cite{icecube_limits}. The short-duration point-source sensitivities of the experiments GRAND-200k (red) \cite{Guepin:2022qpl} and POEMMA (pink) \cite{POEMMA} are also plotted.}
  \label{fig:short_duration}
\end{figure}

To estimate the peak sensitivity of BEACON to short-duration transients ($t<1000 \, \text{s}$), we insert the maximum instantaneous effective area at each energy into Eq. \ref{eq:sens}. The result for 100 stations is depicted by the solid, black curve in figure \ref{fig:short_duration}. The 1000-station estimate is shown with the dashed, black curve. The sensitivity curves of BEACON are compared to the predicted all-flavor neutrino fluence from the extended (blue) and prompt (purple) emission of a short gamma ray burst, viewed on axis, at a distance of 40 Mpc \cite{KMMK}, as well as the upper limits on the neutrino spectral fluence from GW170817, during a $\pm500$ s window, imposed by IceCube and the Pierre Auger Observatory \cite{icecube_limits} . Additionally, the short-duration point-source sensitivities of the planned Earth-skimming tau neutrino detectors GRAND-200k (red) \cite{Guepin:2022qpl} and POEMMA (pink) \cite{POEMMA} are shown.

\begin{figure}[tbp]
  \centering
  \includegraphics[width=0.75\textwidth]{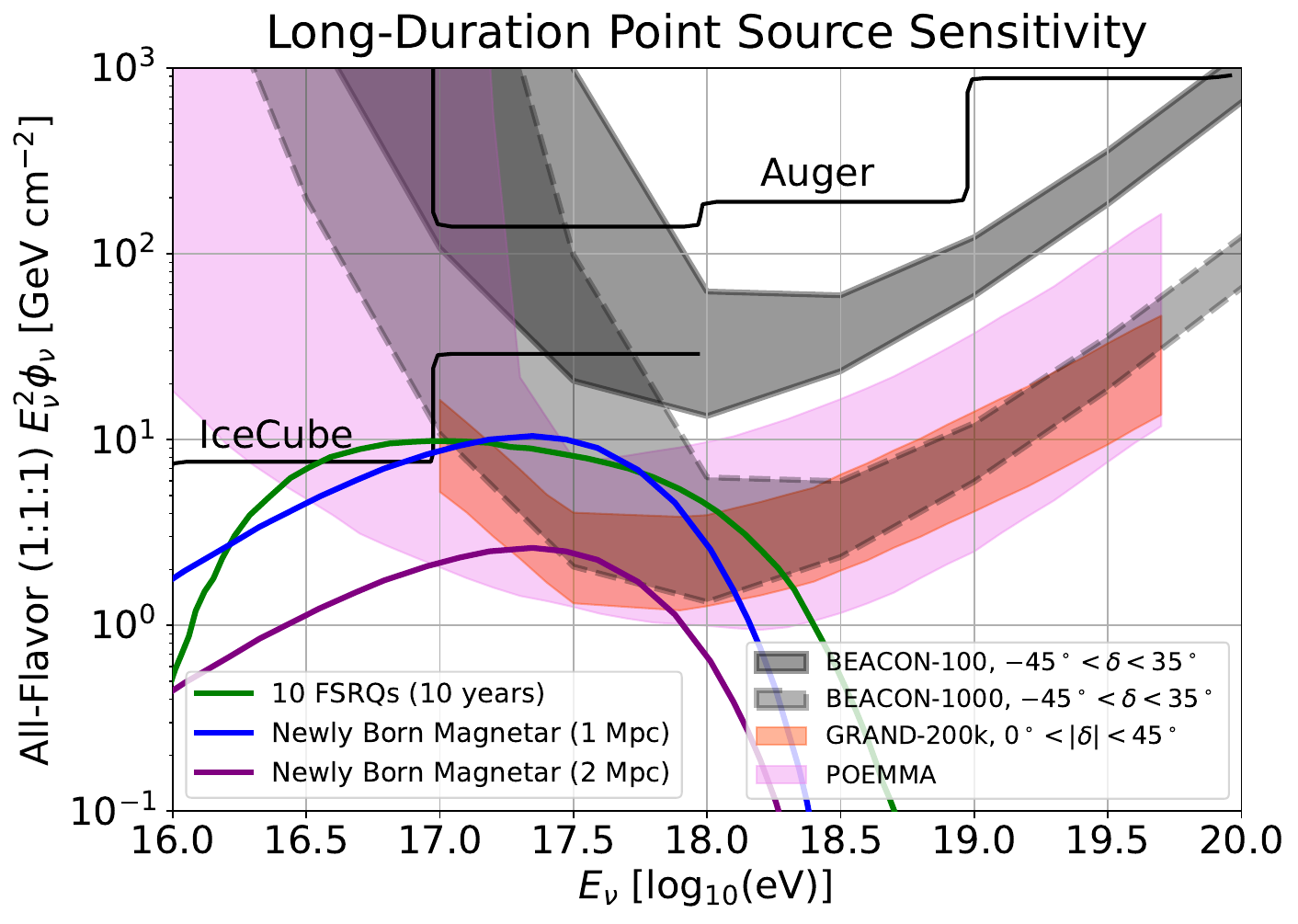}
  \caption{The sensitivity of BEACON to a long-duration ($t> 1 \, \text{day}$) neutrino fluence from a point-source. The result for 100 stations (gray, solid) is obtained by inserting the minimum and maximum day-average effective areas within $-45^\circ<\delta<35^\circ$ into eq. \ref{eq:sens}. The result for 1000 stations (gray, dashed) is predicted by scaling the 100 station result by a factor of 10. These curves are compared to long-duration transient fluence models as well as upper limits imposed by IceCube and the Pierre Auger Observatory (see text) \cite{icecube_limits}. The long-duration point-source sensitivities of the experiments GRAND-200k (red) \cite{GRAND} and POEMMA (pink) \cite{POEMMA} are also plotted.}
  \label{fig:long_duration}
\end{figure}

The sensitivity of BEACON to long-duration transients ($t > 1 \, \text{day}$) is obtained using the day-average effective area. The minimum and maximum day-average effective areas within $-45^\circ<\delta<35^\circ$ are inserted into eq. \ref{eq:sens}, and the 100-station result is depicted by the gray band in figure \ref{fig:long_duration}. The 1000-station estimate is depicted by the dashed gray band. These bands are compared to the stacked all-flavor neutrino fluence resulting from 10 flat-spectrum radio quasars (FSRQs) over a 10-year period \cite{Guepin:2022qpl, Oikonomou:2021akf}, the 2-day all-flavor neutrino fluence resulting from a newly born magnetar at a distance of 1 (blue) and 2 (purple) Mpc \cite{Murase:2009pg}, and the upper limits on the neutrino spectral fluence from GW170817, during a 2-week window, imposed by IceCube and the Pierre Auger Observatory \cite{icecube_limits}. Also shown is the long-duration point-source sensitivity of GRAND-200k (red) \cite{GRAND} and POEMMA (pink) \cite{POEMMA}.

\begin{figure}[tbp]
  \centering
  \includegraphics[width=1.0\textwidth]{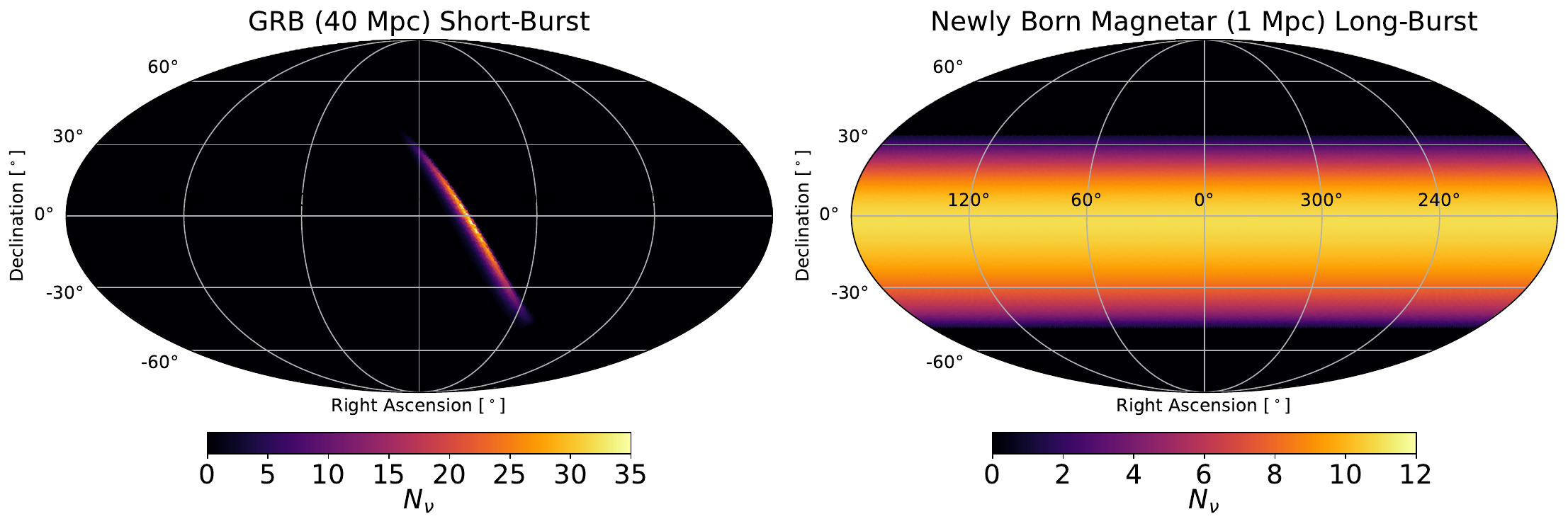}
  \caption{The predicted number of neutrino detections by a 1000-station BEACON in equatorial coordinates for a short-duration transient, the extended emission of a sGRB at 40 Mpc \cite{KMMK} (left), and a long-duration transient, a newly born magnetar at 1 Mpc \cite{Murase:2009pg} (right).}
  \label{fig:num_events}
\end{figure}

The expected number of neutrino detections from an astrophysical source located at ($\alpha, \delta$) is given by
\begin{equation}
   N  = 
   \int \phi_{\nu_\tau}(E_\nu) \, A(E_\nu, \alpha,\delta) \: dE_\nu,
   \label{eq:number}
\end{equation}
where $\phi_{\nu_\tau}(E_\nu)$ is the tau neutrino fluence from the source in units of $(\text{GeV cm}^2)^{-1}$. In figure \ref{fig:num_events}, on the left, we use the instantaneous effective area of a 1000-station BEACON to predict the number of neutrino detections resulting from a short-duration transient, specifically the extended emission of a sGRB from ref. \cite{KMMK} at a distance of 40 Mpc. The maximum number of neutrino detections is 35. The maximum distance in which a sGRB produces at least one detected neutrino is 237 Mpc. On the right, we use the day-average effective area to determine the same result for a long-duration transient, specifically the two-day neutrino emission of a newly-born magnetar, as predicted in ref. \cite{Murase:2009pg}, at a distance of 1 Mpc. Here, the maximum number of detections is 11. The farthest distance in which a newly-born magnetar produces at least one detected neutrino is 3.4 Mpc.

\subsection{Sensitivity to Diffuse Fluxes}

\begin{figure}[tbp]
  \centering
  \includegraphics[width=0.6\textwidth]{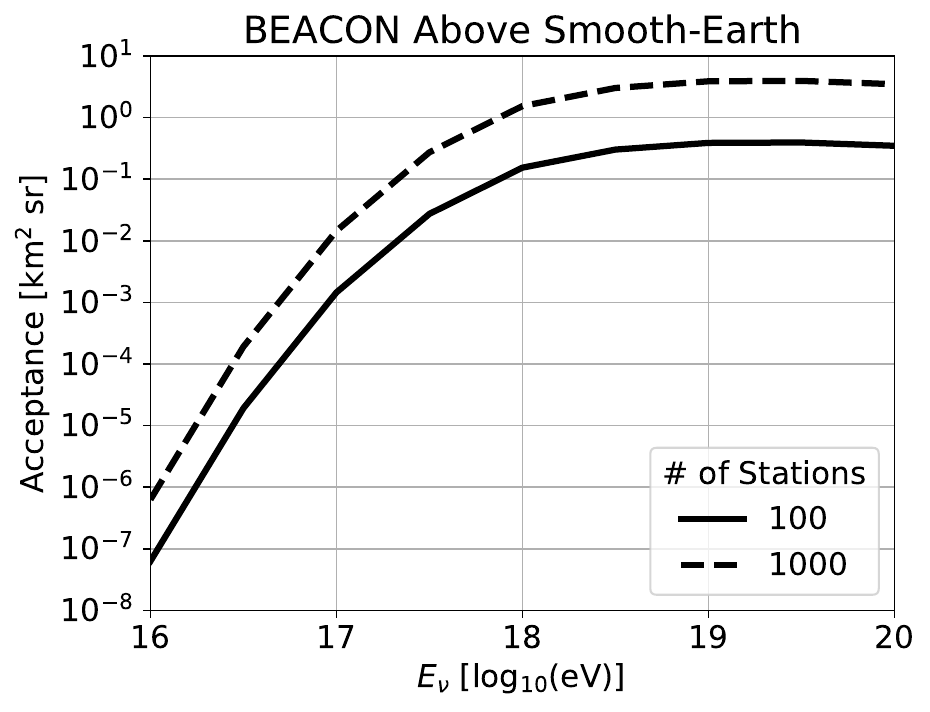}
  \caption{The acceptance of 100 and 1000 BEACON stations at an altitude of 3 km, spaced 3 km apart. Calculated using Eq. \ref{eq:acceptance} and $A(E_\nu,\alpha,\delta)$ as determined by MARMOTS.}
  \label{fig:acceptance}
\end{figure}

The sensitivity of a detector to an all-flavor $dN_\nu/d{E}_\nu \propto E_\nu^{-2}$ diffuse neutrino flux is given by
\begin{equation}
   E_\nu^2 \Phi_\nu  = 
   \frac{2.44} {\text{ln}(10)} \frac{3 \: E_\nu}{\left<A\Omega\right>T}, \label{eq:diffuse_sens}
\end{equation}
where $\left<A\Omega\right>$ is the acceptance and T is the detector live-time. The acceptance is a function of neutrino energy and can be found by integrating each $A(E_\nu,\alpha,\delta)$ over right ascension and declination:
\begin{equation}
   \left<A\Omega\right>(E_\nu)  = 
   \int_0^{2\pi} \int_{-\pi/2}^{\pi/2} A(E_\nu, \alpha, \delta) \cos\delta \; d\delta \, d\alpha. \label{eq:acceptance}
\end{equation}
The acceptance of 100 and 1000 BEACON stations is shown in figure \ref{fig:acceptance}.

\begin{figure}[htbp]
  \centering
  \includegraphics[width=0.8\textwidth]{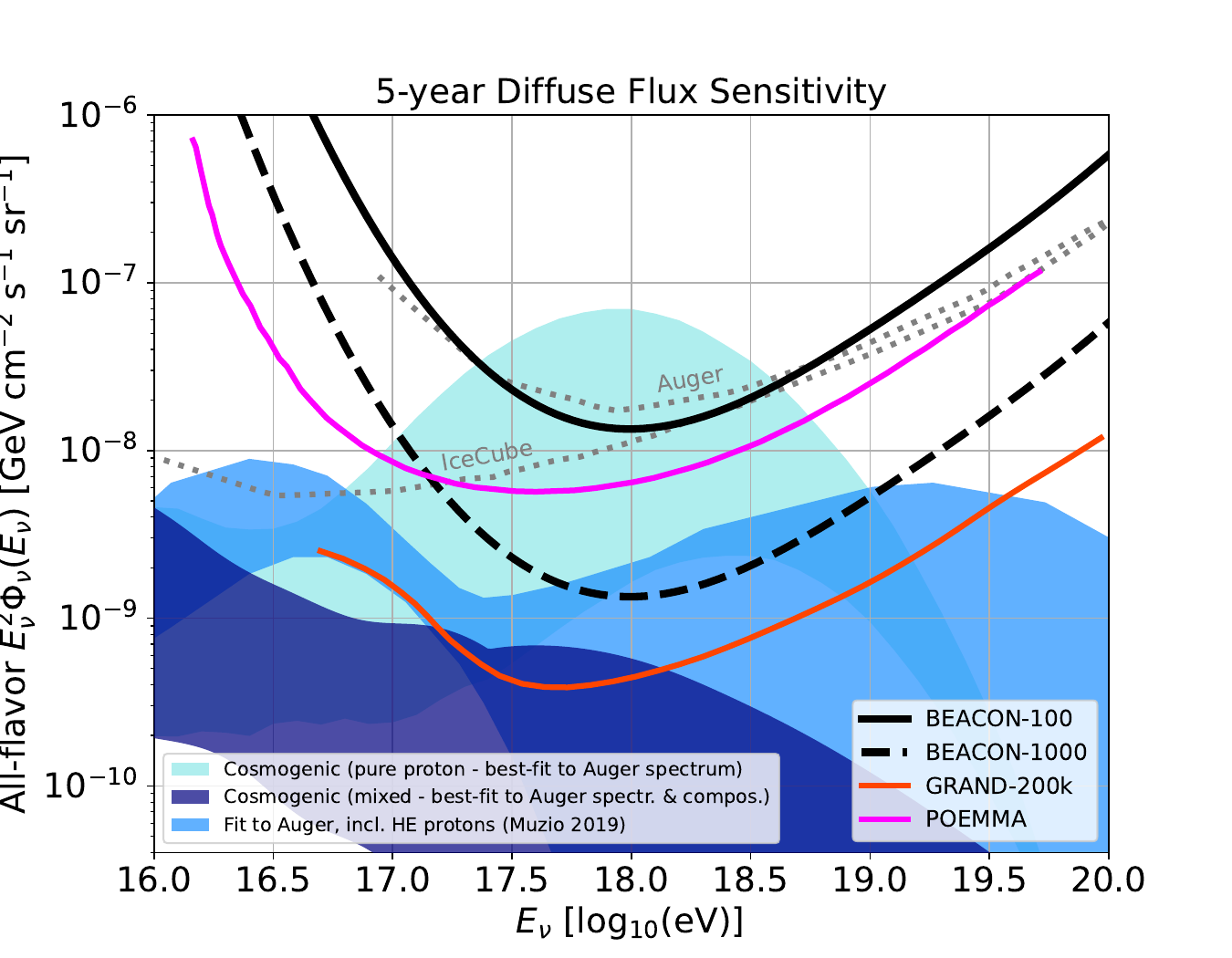}
  \caption{The sensitivity of 100 (black, solid) and 1000 (black, dashed) stations of BEACON to an all-flavor $dN_\nu/dE_\nu \propto E_\nu^{-2}$ diffuse-flux of neutrinos, over a 5-year exposure. These curves are compared to diffuse neutrino flux models \cite{vanVliet:2019cpl, PierreAuger:2022atd, Muzio:2019leu}, as well as the all-flavor upper limits imposed by IceCube \cite{IceCube:2025ezc} and the Pierre Auger Observatory \cite{AbdulHalim:2023SN}. Also shown is the diffuse flux sensitivity of GRAND-200k (red) \cite{GRAND} and POEMMA (pink) \cite{POEMMA:2020ykm}.}
  \label{fig:diffuse}
\end{figure}

Using these acceptances, and Eq. \ref{eq:diffuse_sens}, we can predict the 5-year diffuse flux sensitivity of 100 and 1000 stations of BEACON. This is shown in figure \ref{fig:diffuse} via the solid and dashed black curves respectively. These curves are compared to diffuse neutrino flux models \cite{vanVliet:2019cpl, PierreAuger:2022atd, Muzio:2019leu}, as well as the all-flavor upper limits imposed by IceCube \cite{IceCube:2025ezc} and the Pierre Auger Observatory \cite{AbdulHalim:2023SN}. Also shown is the 5-year diffuse flux sensitivity of GRAND-200k (red) \cite{GRAND} and POEMMA (pink) \cite{POEMMA:2020ykm}. BEACON-100 will match existing limits set by IceCube and Auger at 1 EeV, while BEACON-1000 will be able to probe a variety of cosmogenic flux models.

\section{The Effects of Topography}
\label{sec:topo}

The results up till now have used the geometry described in section \ref{sec:geom}: the Earth is a smooth sphere, above which the stations float at their specified locations. This model is an approximation for a mountaintop detector overlooking a valley at sea-level. In reality, the topography around the mountaintop stations will likely be much more complex. Topography can increase effective area by providing a larger surface area in view of the stations (increasing $A_g$), as well as by providing a larger interaction volume for tau neutrinos (increasing $P_\text{exit}$). On the other hand, topography may decrease effective area by blocking line-of-sight to the resulting EAS (decreasing $P_\text{detect}$). Which effect proves dominant is ultimately site-dependent and requires simulation to resolve. As a result, we have created a branch of MARMOTS which accounts for topography. 

\subsection{Incorporating Topography into the Monte Carlo}

\begin{figure}[tbp]
 \centering
 \subfloat{{\includegraphics[trim={1.8cm 1.5cm 0.2cm 2.3cm},clip, width=0.58\textwidth]{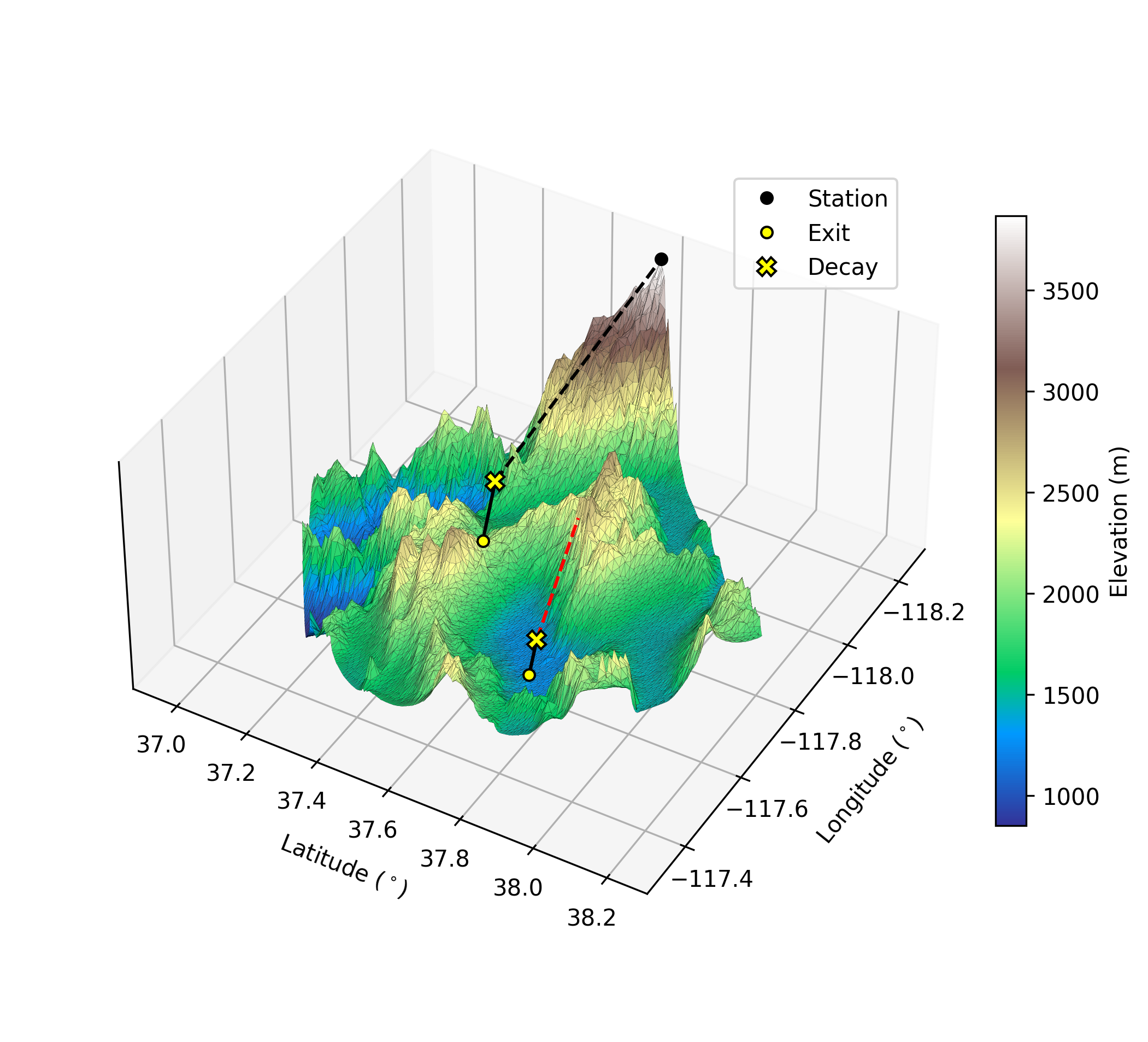}}}
 \,
 \subfloat{{\includegraphics[width=0.40\textwidth]{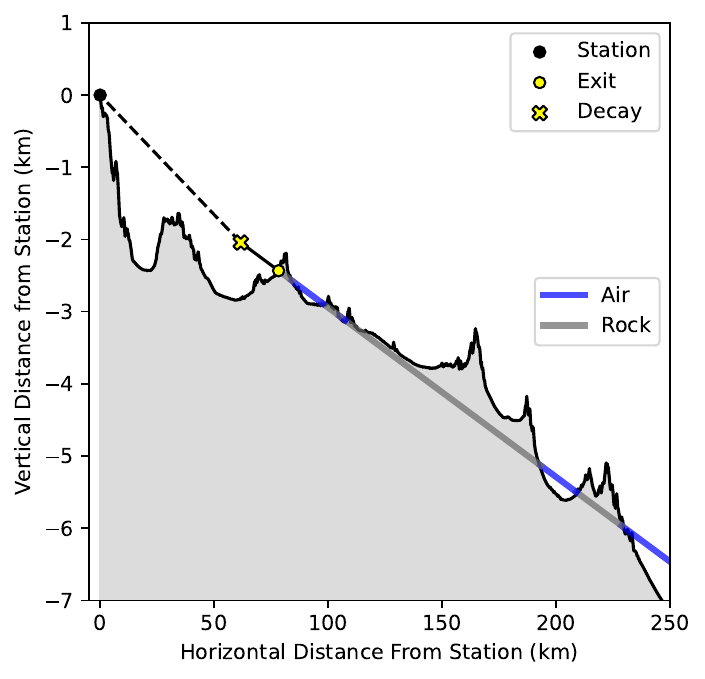}}}
   \caption{Left: Triangulated mesh of the Earth's surface within the BEACON prototype's field-of-view, out to a distance of 80 km. Two tau lepton exit and subsequent decay locations are shown, one of which has line-of-sight to the station and another which does not. Right: 2D slice of the Earth along the propagation axis of an Earth-skimming event. $P_{i,\text{exit}}$ is a function of grammage, calculated by integrating density backwards from the exit point. By finding intersections between the particle axis and the topographic mesh, we can determine when the particle is traversing air instead of rock. From the decay point, line-of-sight to the station is determined.}
  \label{fig:mesh}
\end{figure}

From each station, the location of an arc spanning the FoV, at a specified distance away, is found. The station and this arc form a circular sector on the surface of the Earth. This sector must be large enough such that it contains all potentially detectable tau-lepton exit points for a given station. The union of all the sectors is found and then triangulated, similar to what was done with the cone-intersection areas in \ref{sec:geom}. At every triangle vertex, the elevation corresponding to that latitude and longitude is interpolated from the Shuttle Radar Topography Mission (SRTM) database \cite{SRTM}. The SRTM database is a topographic database containing elevation data at a resolution of one arcsecond (30 m) for nearly 80$\%$ of the Earth. Access to the SRTM is provided for free by NASA. By setting the elevation of each triangle vertex in this way, we have thus created a triangulated mesh of the Earth's surface within the FoV of all the stations. An example of a surface mesh for a single station is shown on the left in figure \ref{fig:mesh}.

To determine $A_g$ for a given source direction, the angles between $\hat{r}$ and the vectors extending from the center of each triangle to each station are first calculated. Triangles with angles less than $\theta_{max}$ are kept, as they may contain detectable tau lepton exit points. $A_g$ is then the sum of the areas of these remaining triangles. Tau lepton exit points are then uniformly sampled within $A_g$, like before. The dot product $(\hat{r}_i\cdot\hat{u}_i)$ in Eq. \ref{eq:eff_area} is replaced with the dot product $(\hat{r}_i\cdot\hat{n}_i)$, where $\hat{n}_i$ is the vector normal to the triangle on which the exit point was sampled.

$P_{i,\text{exit}}$ is a function of grammage, the amount of matter traversed by the tau neutrino and subsequent tau lepton before the exit point. The density of Earth as a function of depth is modeled using the Preliminary Reference Earth Model (PREM) \cite{DZIEWONSKI1981297}. For steeply up-going tau leptons, the grammage is calculated by finding the cord length traversed through a spherical Earth, and then integrating the Earth's density along it. For Earth-skimming events, intersections between the topographic mesh and the backwards particle axis are found. These intersections are used to determine when air (negligible contribution to grammage) is being traversed rather than ground. This is depicted on the right in figure \ref{fig:mesh}. The total grammage is found by integrating density along the backwards particle axis. $P_{i,\text{exit}}$, given the total grammage traversed, is then determined via interpolation of the NuTauSim lookup tables described in section \ref{sec:exit}.

Tau lepton decays are sampled the same as before, however topography may now block a station's line-of-sight to a given decay. For each station, the line-of-sight to each tau decay is checked. This is done by once again finding intersections between the particle axis (now forward) and the topographic mesh. If an intersection is present then the line-of-sight between a station and a tau decay is interrupted by topography, and the event is assumed to be undetectable. A tau decay with and without line-of-sight is depicted in figure \ref{fig:mesh}. The resulting SNR from the remaining in-view events is found, and $P_{i,\text{detect}}$ is determined.

\subsection{Results}

To illustrate the potential impact topography can have on effective area, we calculate the effective area for a single 10-antenna station located at the BEACON prototype site, with and without topography. The prototype site is located within the White Mountains of California (latitude: $37.589^\circ$ N, longitude: $118.238^\circ$ W, altitude: 3.85 km) and overlooks Fish Lake Valley to the East. 

\begin{figure}[tbp]
  \centering
  \includegraphics[width=1.0\textwidth]{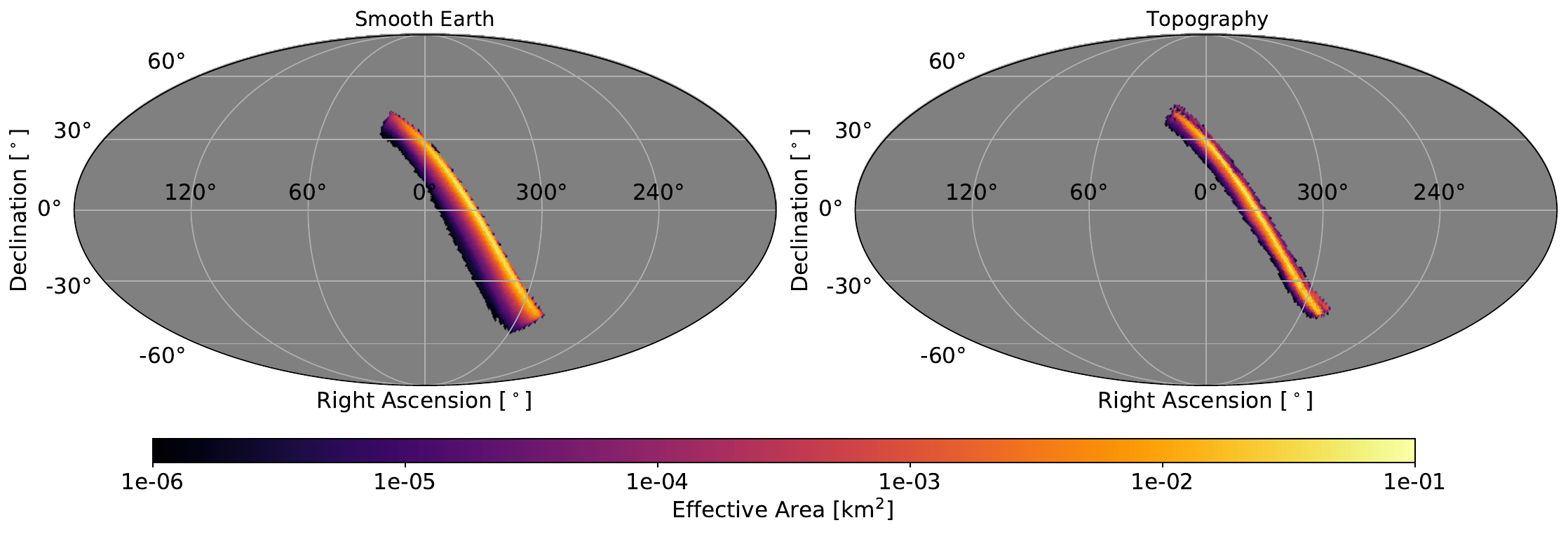}
  \caption{The instantaneous effective area ($E_\nu=1$ EeV) for a single 10-antenna station located at the BEACON prototype site (latitude: $37.589^\circ$ N, longitude: $118.238^\circ$ W, altitude: 3.85 km) using the smooth Earth version of MARMOTS (left) and the topographic version (right).}
  \label{fig:topo_aeff}
\end{figure}

The instantaneous effective area at $E_\nu=1$ EeV for this arrangement, with and without topography, is shown in figure \ref{fig:topo_aeff}. There is a $\sim 15\%$ increase in the maximum effective area when including topography. This seems to be driven primarily by the increase in geometric area within view of the station. Additionally, we gain some sensitivity just above the horizon, likely due to tall topographic features near the horizon. On the other hand, the effective area at angles far below the horizon is greatly diminished, shrinking instantaneous sky-coverage. This appears to be due to nearby topographic features blocking line-of-sight to tau decays at steep elevation angles.

\begin{figure}[tbp]
  \centering
  \includegraphics[width=1.0\textwidth]{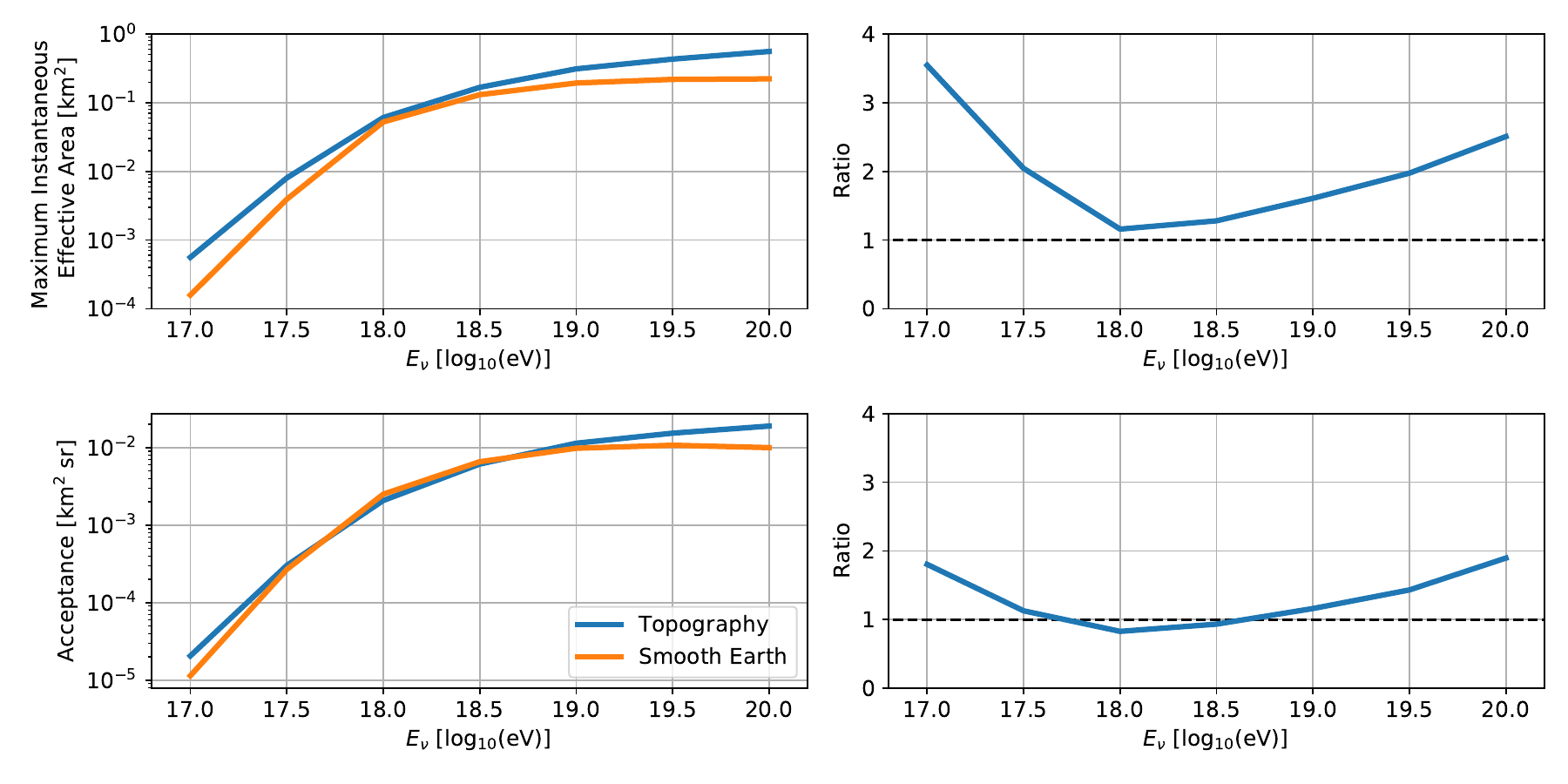}
  \caption{The maximum instantaneous effective area (top left) and the acceptance (bottom left), as a function of energy, for a single 10-antenna station located at the BEACON prototype site (latitude: $37.589^\circ$ N, longitude: $118.238^\circ$ W, altitude: 3.85 km) using the smooth Earth version of MARMOTS (orange) and the topographic version (blue). The ratio of the topographic results to the smooth Earth results are shown on the right.}
  \label{fig:ratio}
\end{figure}

Figure \ref{fig:ratio} demonstrates how the maximum instantaneous effective area and the acceptance are impacted by the inclusion of topography as a function of neutrino energy. Including topography increases the maximum instantaneous effective area at all energies, however the effect is strongest at the extremes. The increase at lower energies is likely due to the exiting tau leptons being brought closer to the station, as the valley is now no longer at sea-level, while the increase at higher energies is likely due to the overall increase in surface area when looking out towards the horizon. Topography has a similar energy-dependent effect on acceptance, however it is less pronounced. This is likely due to the reduction in instantaneous sky-coverage resulting from line-of-sight blockage. Acceptance is integrated across the sky, so an increase to the maximum effective area may be unable to compensate for a decrease in sky-coverage. This is most easily seen at $10^{18}$ eV, where the acceptance with topography is slightly lower than that without topography, despite an increase to the maximum effective area.

For this specific site, the location of the BEACON prototype, including topography improves our maximum instantaneous effective area (and thus our sensitivity to short-duration transients) by over a factor of 3 below $10^{17}$ eV. This energy range is particularly relevant for the detection of astrophysical transients, as this is where most transient flux models peak \cite{Ackermann:2019cxh, Guepin:2022qpl}. It is possible that for other sites the inclusion of topography has a different effect, either positive or negative, due to the presence of different topographical features. The example of the prototype site shown here indicates that topography likely impacts the effective area by factors of a few. MARMOTS can be used to identify the most optimal sites for future BEACON stations, and once they have been chosen, an updated sensitivity estimate can be provided.

\section{Discussion}
\label{sec:discussion}

% Points
% BEACON is sensitive to point sources and transients with deep effective area in narrow field of view
% somewhat efficcient
% Particularly sensitive to high rate transients
% Does this with a design optimized for high energies using a high altitude detector and phased arrays
% One advantage is triggered transient / point source searches where we can tune the trigger threshold

By observing the horizon with phased arrays from elevated sites, BEACON achieves a very large instantaneous effective area to ultra-high energy tau neutrinos. This in turn provides excellent sensitivity to short-duration transients, as shown in figure \ref{fig:short_duration}. With just 100 stations, BEACON has the potential to detect nearby short gamma-ray bursts. While the instantaneous effective area is quite large, it exists only within a narrow field-of-view. The percentage of the sky containing a detectable sGRB at 40 Mpc, for example, is just $\sim2.4\%$ (see figure \ref{fig:num_events}). This deep but narrow instantaneous effective area means that while the rate of transients occurring within the field-of-view is reduced, those which do occur are more likely to be detected. Fermi GBM detects $\sim240$ GRBs per year, about 40 of which are short GRBs \cite{Fermi}. For a FoV of $2.4\%$, BEACON should therefore observe about 5 long and 1 short GRBs per year. Sky-coverage could be increased by splitting up BEACON along multiple mountain ranges, however the peak sensitivity would then be decreased assuming the total number of stations remains constant.

Except at the poles, BEACON must be pointed either East or West in order to best utilize the geomagnetic effect. This leads to a day-average effective area with a large field-of-view ($\sim70\%$ of the sky), but a maximum effective area which is greatly reduced relative to the instantaneous case (by about two orders of magnitude). As a result, the opposite becomes true for long-duration transients: the rate of occurrences within the field-of-view is enhanced, while the likelihood to detect any given occurrence is reduced. Nevertheless, with 1000 stations BEACON can still effectively probe various long-duration transient fluence models as shown in figure \ref{fig:long_duration}. The large daily field-of-view will also enable more follow-ups to multi-messenger alerts.

BEACON utilizes phased arrays which enables an interesting capability for multi-messenger astronomy. The locations and number of beams for a phased-array can be adjusted digitally, and their trigger thresholds individually tuned. Therefore, if a multi-messenger alert is received, BEACON can point one of its beams in the direction of the source. It can then increase the threshold of the other beams or disable them entirely. This in turn will allow the threshold of the relevant beam to be lowered, increasing the neutrino sensitivity in that particular direction. Lowering the trigger threshold particularly improves the sensitivity below 1 EeV, an important energy regime for transient detection.

\begin{figure}[tbp]
 \centering
 \subfloat{\includegraphics[width=0.49\textwidth]{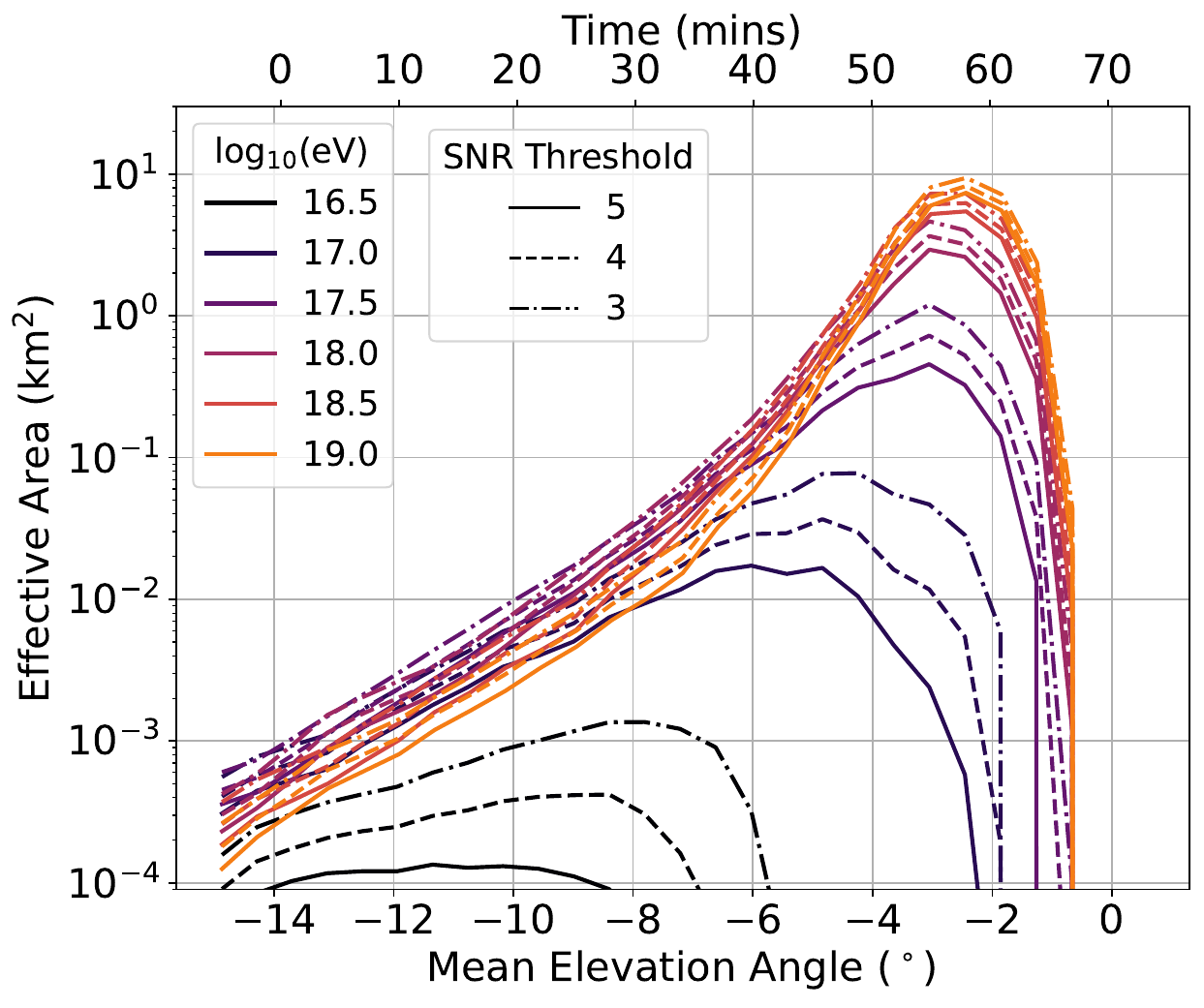}}
 \,
 \subfloat{\includegraphics[width=0.49\textwidth]{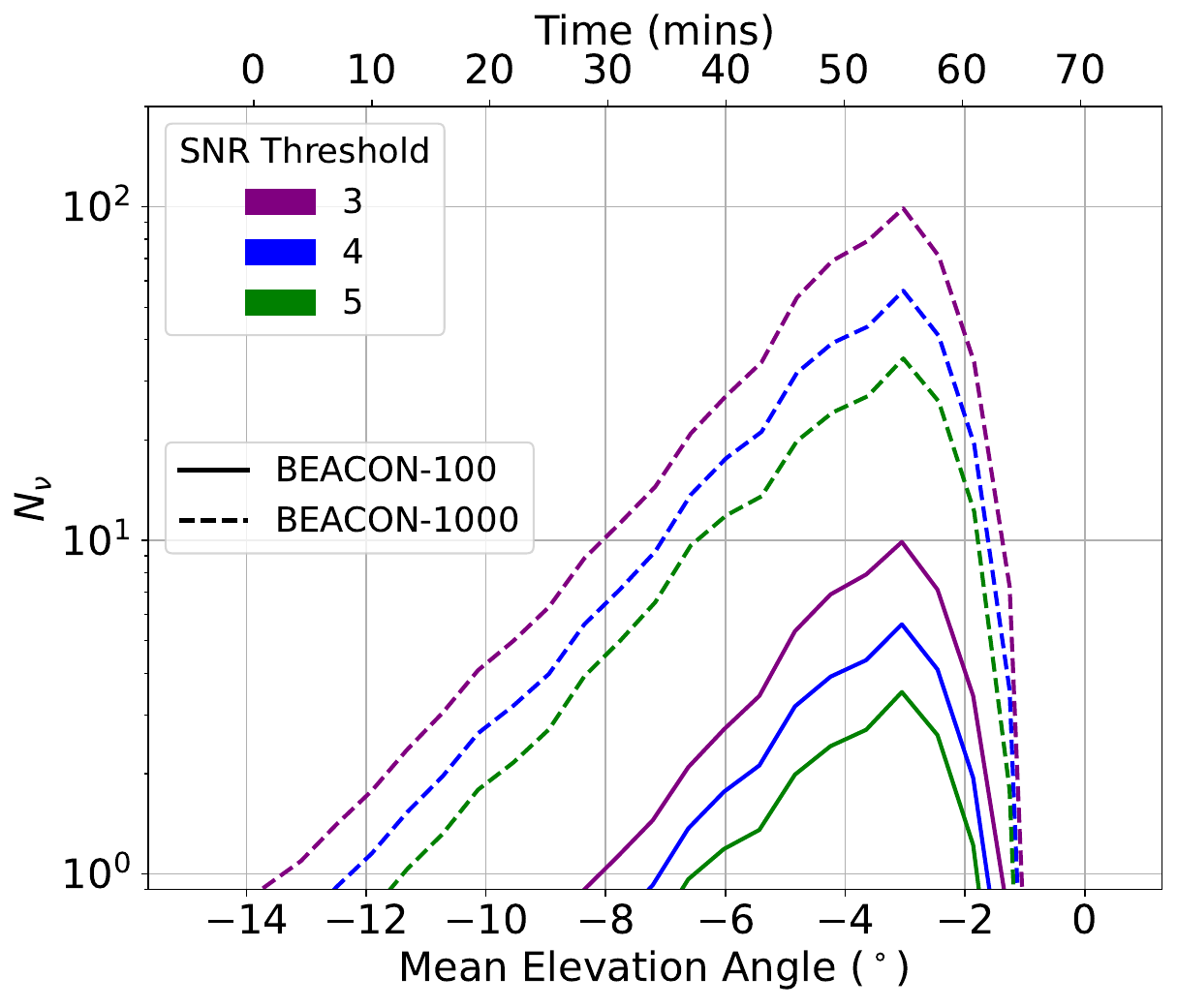}}
  \caption{Left: Effective area of 100 BEACON stations as a function of mean elevation angle (or equivalently time), at various neutrino energies and trigger thresholds. At $t=0$ the source ($\alpha = 0^\circ$, $\delta=0^\circ$) is located approximately $14^\circ$ below horizontal for each station. As time progresses, the Earth rotates and the source rises towards and then above the horizon. Right: Number of detected neutrinos for a sGRB occurring at time $t$, at a distance of 40 Mpc, given the effective areas on the left. Results are shown for three SNR thresholds, and for 100 and 1000 stations.}
  \label{fig:source_over_time}
\end{figure}

The effect of changing the trigger threshold is demonstrated in Figure \ref{fig:source_over_time}. Shown on the left is the effective area of 100 BEACON stations to a single point-source as a function of time, or equivalently, the average source elevation angle. The effective area is plotted for different neutrino energies and for three different SNR trigger thresholds: 5 (normal operation), 4, and 3. At $t=0$, the source (located at $\alpha=0^\circ , \delta=0^\circ$) sits $\sim14^\circ$ below the horizontal of each station. As time progresses, the Earth rotates, bringing the source closer to the horizon and thus changing the effective area in its direction. On the right, these effective areas are translated into how many neutrinos may be detected if this source were an sGRB that occurred at time $t$ (the extended emission of a sGRB lasts only $\sim100$ s \cite{Kisaka:2015mza}). We see that with 100 (1000) stations, there is a roughly 30-minute (1-hour) window in which a nearby sGRB occurring at a particular source location is detectable by BEACON. The duration of this time window increases when the trigger threshold is lowered. The maximum number of neutrinos is detected if the sGRB occurs when viewed just below the horizon (elevation angle of $\sim-3^\circ$). With SNR thresholds of 5, 4, and 3 the maximum number of detected neutrinos with BEACON-100 (BEACON-1000) is 3 (35), 5 (56), and 9 (98). Lowering the threshold in response to a multi-messenger alert can therefore greatly increase the likelihood and significance of a detection.

The short time windows in which many transients remain detectable imposes a latency requirement on BEACON. In the example above, BEACON must be able to receive a multi-messenger alert, redirect its beams, and adjust the trigger thresholds within a $\sim100$ s time window. Beam directions and trigger thresholds are programmed digitally, and can therefore be adjusted automatically. The speed at which multi-messenger alerts are communicated is also continuously improving. In the case of GRB 160821B, MAGIC was able to receive an alert and slew to target in only 24 seconds after the initial burst detection by Swift \cite{MAGIC:2020ikk}. 

By utilizing high elevation sites and phased arrays, BEACON achieves sensitivity to ultra-high energy neutrinos in an efficient manner. BEACON-100 has a short-duration transient sensitivity comparable to other Earth-skimming tau neutrino detectors while employing only 1,000 dual-polarized, short-dipole antennas in total. These antennas are relatively inexpensive to construct and deploy. BEACON-1000 (10,000 antennas) similarly achieves a long-duration transient sensitivity and diffuse flux sensitivity comparable to other proposed experiments, improving upon existing limits by about a factor of 10. The BEACON prototype has achieved a pointing resolution of $\sim1^\circ$ using only four phased antennas \cite{Southall:2022yil}. An array of 10 antennas will thus likely achieve a sub-degree pointing resolution, aiding BEACON in both issuing and following-up multi-messenger alerts.

The sensitivities predicted here are an approximation using evenly-spaced stations placed along a line of longitude, elevated over a smooth Earth. This is meant to resemble stations placed uniformly along a single North-South mountain range, overlooking a valley to the East. As shown, it is possible that local site effects such as topography, or even the strength of the local geomagnetic field, may change the predicted sensitivities.

\section{Conclusions}
% Retierate the whole paper succintly
% Include estimate of FOV
% Incldue summary of how this paper improves on prior design paper
% End noting that BEACON is complementary to other UHE neutrino experiments but emphasis strengths
\label{sec:conclusion}

When fully realized, BEACON will search for Earth-skimming tau neutrinos using mountaintop phased antenna arrays. The beam-formed trigger and high elevation vantage points together achieve a large effective area to tau neutrinos.  This configuration results in a deep sensitivity near the horizon, further enhanced by a directionally-tunable trigger threshold. The sensitivity is particularly optimized for short-duration bursts of neutrinos which, when combined with multi-messenger alerts from new and upcoming wide-field transient instruments such as Vera Rubin, can enhance the likelihood and significance of discovery. 

This work emphasizes the importance of a large instantaneous effective area in searching for short-duration bursts of neutrinos. Short GRBs occurring within the field-of-view of only 100 stations of BEACON would be visible on a case-by-case basis. Neutrino bursts lasting a day or more require additional stations to be readily visible, however with 100 stations may still be detectable using stacked searches. As was found in our prior study, the sensitivity of BEACON-1000 to the diffuse flux is comparable to other proposed air shower experiments while using a modest number of antennas. Lastly, we find that local topography effects can further improve the effective area, improving the sensitivity at some energies by factors of a few.

We note that the sensitivity of BEACON depends on the exact arrangement and locations of the stations. In this study, for simplicity, we arranged 100 stations along a North-South line centered on the prototype site, with stations spaced 3 km apart and elevated 3 km above sea level. Each station consisted of 10 phased antennas, faced East, and had a $120^\circ$ field-of-view. This setup was chosen to emulate stations placed along the side of a single mountain range running North-South, such as the Sierra Nevada or Andes Mountains. One could instead distribute BEACON along multiple mountain ranges, however. This would create multiple bands of effective area across the sky, instead of the single one shown in figure \ref{fig:effective_area}. Doing this will therefore increase instantaneous sky-coverage, however with the total number of stations constant, the effective area within each band will be lower and the point-source sensitivity thus reduced. One must therefore decide if it is better to have a deep effective area within a narrow region of the sky, or a shallower effective area across a larger portion of the sky. Deep and narrow instruments are better suited for detecting neutrinos from known source classes, while wide and shallow instruments are better suited for serendipitous discovery \cite{Kotera:2025jca}.

The Monte Carlo MARMOTS was developed to determine the point-source effective area of any configuration of elevated phased arrays. This is an improvement over our previous simulation which only calculated the isotropic acceptance. Additionally, MARMOTS properly accounts for overlapping individual effective areas when calculating the total effective area. MARMOTS also includes improved electric field simulations from ZHAireS-RASPASS, as well as a more realistic antenna model. An alternate branch of MARMOTS has also been developed which properly accounts for local topography which, as demonstrated here, has an effect on neutrino sensitivity and should therefore be accounted for when selecting experiment sites.

To demonstrate triggering on EAS using an elevated phased array, the BEACON prototype will first be used to detect cosmic rays \cite{Southall:2022yil}. Down-going EAS initiated by cosmic rays emit a radio impulse due to the same geomagnetic mechanism as up-going EAS from tau leptons, while also having a greater flux. The sensitivity of the prototype to cosmic rays has been predicted using similar electric field simulations, as well as the same detector model, as that used in MARMOTS \cite{Zeolla:2021cbb}. A validation of the prototype's cosmic ray sensitivity will thus serve as an important step in validating the neutrino sensitivity predicted here. A search to identify cosmic rays within the data is currently ongoing.

\appendix
\section{Instantaneous Effective Area}
\label{sec:appendix}
%Please always give a title also for appendices.

\begin{figure}[htb]
  \centering
  \includegraphics[width=1.0\textwidth]{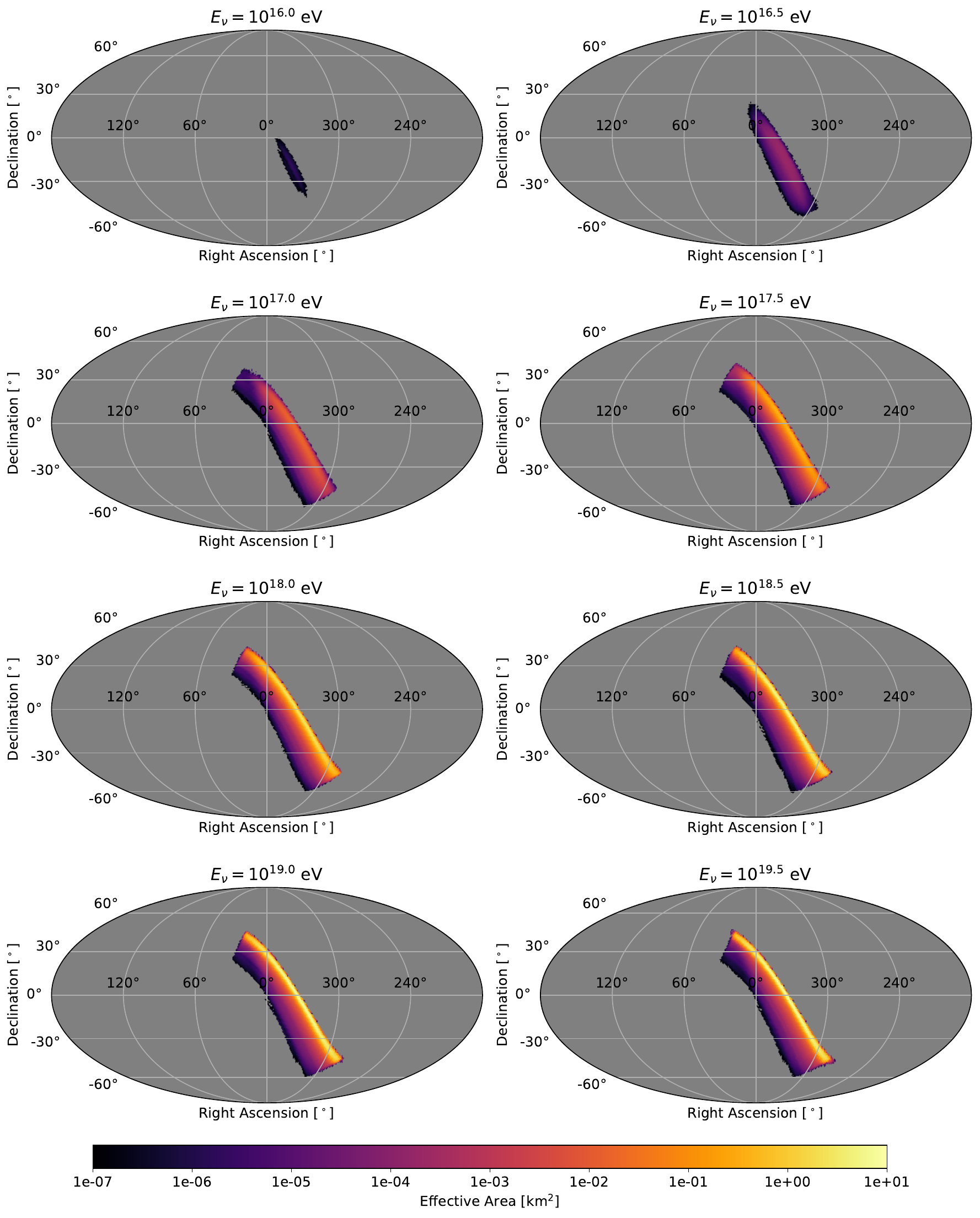}
  \caption{Mollweide projection of the instantaneous effective area, as a function of right ascension and declination, for a 100-station BEACON at a series of neutrino energies (Greenwich Sidereal Time of 0:00).}
  \label{fig:effective_areas}
\end{figure}

Shown in figure \ref{fig:effective_areas} are Mollweide projections of the instantaneous point-source effective area, for the arrangement of 100 BEACON stations described in section \ref{sec:results}, at neutrino energies ranging from $10^{16}$ to $10^{19.5}$ eV in half-decade steps. The change in the instantaneous effective area as neutrino energy is increased is primarily due to an increase in:
i) the probability of $\nu_\tau$ interactions within the Earth,
ii) the range of off-axis angles ($\theta_\text{view}$) in which showers are detectable,
and iii) the amplitude of the emitted electric fields, and thus the distance in which a tau-decay is detectable.

\acknowledgments

This work is supported by NSF Awards $\#$ 2033500, 1752922,
1607555, and DGE-1746045 as well as the Sloan Foundation,
the RSCA, the Bill and Linda Frost Fund at the California Polytechnic State University, and NASA (support through JPL and Caltech as well as Award $\#$ 80NSSC18K0231). This work has received financial support from Ministerio de Ciencia, Innovaci\'on y Universidades/Agencia Estatal de Investigaci\'on, MICIU/AEI/10.13039/501100011033, Spain (PID2022-140510NB-I00, PCI2023-145952-2, RYC2019-027017-I, and Mar\'\i a de Maeztu grant CEX2023-001318-M); Xunta de Galicia, Spain (CIGUS Network of Research Centers and Consolidaci\'on 2021 GRC GI-2033 ED431C-2021/22 and 2022 ED431F-2022/15); and Feder Funds. We thank the NSF-funded White Mountain Research Station for their support. Computing resources were provided by the University of Chicago Research Computing Center and the Institute for Computational and Data Sciences at Penn State.

%\paragraph{Note added.} This is also a good position for notes added
%after the paper has been written.

% Bibliography

%% [A] Recommended: using JHEP.bst file
\bibliographystyle{JHEP}
\bibliography{biblio.bib}

\end{document}